\documentclass[12pt,letterpaper]{article}

\usepackage{times}
\usepackage{t1enc}
\usepackage[latin1]{inputenc}
\usepackage[english]{babel}
\usepackage{graphicx}

\usepackage{bm}
\usepackage{amssymb}
\usepackage{amsfonts}
\usepackage{amsmath}

\DeclareMathOperator{\sign}{sign}
\renewcommand{\baselinestretch}{1.75}

\begin{document}

\title{Regression Models for Directional Data Based on Nonnegative Trigonometric Sums}
\renewcommand{\baselinestretch}{1.00}
\author{Fern\'andez-Dur\'an, Juan Jos\'e and Gregorio-Dom\'inguez, Mar\'ia Mercedes  \\
ITAM\\
E-mail: jfdez@itam.mx}
\date{}
\maketitle

\renewcommand{\baselinestretch}{1.75}

\begin{abstract}
The parameter space of nonnegative trigonometric sums (NNTS) models for circular data is the surface of a hypersphere; thus,
constructing regression models for a cir\-cular-dependent variable using NNTS models can comprise fitting great (small) circles on the
parameter hypersphere that can identify different regions (rotations) along the great (small) circle.
We propose regression models for circular- (angular-) dependent random variables in which the
original circular random variable, which is assumed to be distributed (marginally) as an NNTS model, is transformed into a linear random variable such that common methods for linear regres\-sion can be applied. The usefulness of NNTS models with skewness and multimodality is shown in examples with simulated and real data.
\end{abstract}


\section{Introduction}

\addtolength{\textheight}{.5in}%

The need to analyze circular variables (angles, orientations, or directions) arises in various disciplines, such as the direc\-tions taken by animals in biology, the conformational angles in a molecule in chemistry (e.g., a protein), the characteriza\-tion of
wind patterns in terms of wind direction and speed in environmetrics, the time of onset of a certain disease or times of death of individuals in a certain population in medi\-cal sciences, the position of events that occur on the surface of the Earth (e.g., the epicenter of an earthquake) or the study of paleomagnetism in geology, and the orientation of planets and other
celestial bodies in relation to the Earth at a certain time in astronomy. In all of these examples, when dealing with circular variables,
it is necessary to consider the periodicity of their density functions. The same constraint applies when constructing regression models for a circular
response and a set of explanatory variables of different types (e.g., circular, linear, and categorical). Early developments in circular-circular regression (i.e., a circular response and a single
circular explanatory variable) and circular-linear re\-gres\-sion (i.e., a circular response and only a linear explanato\-ry variable) and their corresponding bivariate correlation co\-ef\-fi\-cients
have been reviewed by Upton and Fingleton (1989), Fisher (1993), Mardia and Jupp (2000), and Jammalamadaka and SenGupta (2001). In addition, Gould (1969) developed regression models for a circular response considering a von Mises density, where its mean was modeled as a linear func\-tion of explanatory variables not satisfying the periodicity constraint for a circular mean.
Fisher and Lee (1992) and Fisher (1993) considered regression models for a circular response variable with a von Mises distribution while model\-ing its mean using a tangent link function of a linear function of the explanatory variables satisfying the periodicity con\-straint. The concentration parameter of the von Mises
dis\-tri\-bu\-tion can be modeled as an exponential function of other linear functions of the explanatory variables.
However, Presnell et al. (1998) criticized the existing regression models based on circular response because they suffered from the non-identifiability of the parameters and complex and mul\-ti\-modal likelihoods, which rendered determining their global maxima challenging. Thus, Presnell et al. (1998) proposed
a spherically projected multivariate linear model to avoid these issues. Consequently, models based on Presnell et al. (1998) were proposed, such as, those of N\'{u}\~{n}ez-Antonio et al. (2011), and Wang and Gelfand (2013). In addition, Lund (1999) criticized the use of least-squares regression in cases where the model  response was circular and proposed the use of the least circular distance instead. Downs and Mardia (2002) constructed circular-circular regression models wherein the regression curve was a M\"obius circular trans\-for\-ma\-tion with errors following a von Mises distribution. Kato et al. (2008) considered a M\"obius transformation with errors following a wrapped Cauchy distribution. SenGupta et al. (2013) developed models for inverse circular-circular regression, where the value of the unobserved ex\-plana\-to\-ry circular variable was predicted by the observed circular re\-sponse. Polsen and Taylor (2015) proposed a common framework for circular-circular regression based on a tangent link function. Further, while Rivest et al. (2016) developed gen\-er\-al circular regression models for animal movements, Kim and SenGupta (2017) recently constructed models for mul\-ti\-vari\-ate-multiple circular regression. Breckling (1989), Fisher (1993), and Pfaff et al. (2016) analyzed the time series of a circular variable. Fisher (1993) proposed the use of a link function (i.e., the inverse of the tangent function) for
constructing circular time-series models. Holzmann et al. (2016) considered the analysis of circular and circular-linear time series. Further, details regarding  spatial and spa\-tiotem\-po\-ral models for circular variables can be found in Fisher (1993),
Jona-Lasinio et al. (2012), Wang and Gelfand (2014), Wang et al. (2015), Mastrantonio et al. (2016), and other cor\-re\-spond\-ing references therein.

Fern\'andez-Dur\'an (2004) constructed density functions for circular random variables based on the use of nonnegative trigonometric sums (NNTS) models that allow
modeling of skewed and multimodal data. The CircNNTSR R package (R Development Core Team, 2012) contains routines to fit NNTS models, which was explained by Fern\'andez-Dur\'an and Gregorio-Dom\'inguez (2016).
The main objective of this study is to consider the construction of regression models, in which the dependent circular random variable is considered to be distributed as an NNTS model.

The remainder of this paper is organized as follows. Sec\-tion 2 explains how to transform the dependent NNTS cir\-cu\-lar random variable to a unique linear
random variable such that commonly used linear regression models can be fit to the transformed linear variable. Section 3 presents examples of the proposed methodology
on real datasets, including re\-gres\-sion and time-series models. In addition, a simulation study is presented. Finally, Section 4 con\-cludes the paper.

\section{Transforming NNTS Distributed Circular Data to Linear Data}

Consider a sample of random vectors $(\Theta_k, X_{k1}, X_{k2}, \ldots, X_{kp})$ for $k = 1, 2, \ldots \\, n$, where $\Theta_k$ is a circular random variable,
and $X_{k1}, X_{k2}, \ldots, X_{kp}$ is a set of explanatory variables for the $k$-th individual in the sample. The explanatory variables could be of any type, qualitative or quantitative, and in any measurement scale.

We assume that the circular random variables $\Theta_1, \Theta_2, \ldots, \Theta_n$ are distributed as NNTS  distributions. Then,
the density function for $\Theta_k$ is defined as follows (Fern\'andez-Dur\'an 2004):
\begin{equation}
f(\theta)=\frac{1}{2\pi}\sum_{k=0}^M\sum_{m=0}^M c^*_k\bar{c}^*_me^{i(k-m)\theta}=\frac{1}{2\pi} \bm{c}^{*H} \bm{e}^* \bm{e}^{*H} \bm{c}^*
\label{NNTS1}
\end{equation}
where $\bm{c}^*=(c^*_0, c^*_1, \ldots, c^*_M)^\top$ is a vector of complex numbers, that correspond to the vector of parameters of the model and, $\bm{c}^{*H}$ is the Hermitian transpose, satisfying the following constraint:
\begin{equation}
\sum_{k=0}^M ||c^*_k||^2 = 1.
\end{equation}
This constraint specifies that the parameter space is a complex hy\-per\-sphere of dimension $M$, $S_{\mathbb{C}}(M)$ in $\mathbb{C}^{M+1}$. Given this constraint, $c_0$ is specified as a nonnegative real number.

The vector $\bm{e}^*$ is the complex vector of trigonometric terms defined as $\bm{e}^* = (1, e^{-i\theta}, e^{-2i\theta}, \ldots, \\ e^{-Mi\theta})^\top$ with $i=\sqrt{-1}$ and $e^{-ki\theta}=\cos(k\theta) - i\sin(k\theta)$ for $k=0, 1, \ldots, M$. The NNTS density $f(\theta)$ is the squared modulus of the product $\bm{c}^{*H}\bm{e}^*$, implying that an angle $\theta$ for which the derived vector of trigonometric terms, $\bm{e}^*$, obtains a larger modulus for $\bm{c}^{*H}\bm{e}^*$, then obtains a larger probability. Thus,
the parameter vector $\bm{c}^*$ can be interpreted as a multidimensional location parameter, where the argument (angle) of each of its components defines
directions that have a larger probability if the modulus of the corresponding component of the parameter vector is larger than those of the other components. Further, the sum of the moduli of the components of the parameter vector is restricted to one. Parameter $M$ acts as a scale and shape parameter that determines
the maximum number of modes of the NNTS and when it increases the NNTS density becomes less smooth and more concentrated around certain angles depending on the arguments and moduli of the elements of the complex parameter vector $\bm{c^*}$. To operate in a real hypersphere of dimension $2M$, $S(2M)$, the complex vectors
$\bm{c}^*$ and $\bm{e}^*$ are transformed into real vectors $\bm{c}$ and $\bm{e}$ as $\bm{c}=(Re(\bm{c}^*),Im(\bm{c}^*))^\top$
and $\bm{e}=(Re(\bm{e}^*),Im(\bm{e}^*))^\top$, where $Re(\bm{c}^*)$ and $Im(\bm{c}^*)$ are the real and imaginary parts of the elements of the complex vector
$\bm{c}^*$. For vector $\bm{e}^*$, $\bm{e}=(1, \cos(\theta), \cos(2\theta), \ldots, \cos(M\theta),-\sin(\theta), -\sin(2\theta), \ldots, -\sin(M\theta))^\top.$ \\ The NNTS density is a special case of an orthogonal series density estimator (see the review by Izenman, 1999), with the particular advantage that the parameter space
is a hypersphere, and its geometry can be exploited by considering great (small) circles as regression curves. The great circles correspond to the geodesics of the hypersphere. Rivest (1999) examined the fit of small circles in a three-dimensional sphere by considering the least squares estimation.

The proposed methodology comprises three steps:
\begin{enumerate}
\item  Expanding the observed values of the circular dependent variable, $\theta_k$ for $k=1, \ldots, n$, into the parameter hy\-per\-sphere, $S(2M)$, using transformation ${\bf \hat{e}}_k \propto (1,e^{-i\theta_k},e^{-2i\theta_k}, \ldots,e^{-Mi\theta_k})^\top$. This transformation in\-cludes the first $M$ trigonometric moments of (minus) $\theta$. The ${\bf \hat{e}}_k$ vectors are normalized to have unit norm.
\item Identifying the great (small) circle that is the best fit for the ${\bf \hat{e}}_k$ vectors in the parameter hypersphere, $S(2M)$, and transforming
the original circular data, $\theta_k$, to linear data, $Y_k$, by considering the position of ${\bf \hat{e}}_k$ relative to the great (small) circle.
\item Fitting a regression model employing the ordinary least squares and an identity link using a dependent variable as the transformed linear variable $Y_k$ to identify significant explanatory variables related to different segments (ro\-ta\-tions) along the fitted great (small) circle.
\end{enumerate}

As reported by Li and Duan (1989), as the link function of the regression model is unknown and the identity link is being used, the intercept term of the regression model is unidentifiable while the other coefficients of the regression are identifiable up to a multiplicative constant. This problem is mitigated in the case of the proposed methodology when obtaining the NNTS predictive densities by considering the intercept term equal to zero owing to the periodic nature of the great (small) circle regression curves on the hypersphere.

The transformation of the original circular data is equiv\-a\-lent to projecting the circular variables into a great (small) circle in the parameter space $S(2M)$, which is similar to kernel projection methods used in statistical learning, such as support vector machines (Hastie et al. 2009).

\subsection{Great-circle (Geodesic) Regression on Unit Hypersphere}

The shortest distance between two points on the unit hy\-per\-sphere $S(2M)$ corresponds to the length of a segment of a great circle (also known as geodesic), which is defined as follows:
\begin{equation}
\bm{\epsilon} = \bm{a}\cos(\phi) + \bm{d}\sin(\phi)
\label{geodesic}
\end{equation}
where $\bm{\epsilon}$ is a point on the geodesic defined by the orthogonal unit vectors $\bm{a}$ and $\bm{d}$ in $S(2M)$, and $\phi$ is an angle taking values in the interval $[-\pi,\pi)$. Note that the geodesic equation is the same for hyperspheres of any dimension. The definition of NNTS density in Equation (\ref{NNTS1})
implies that $\phi \in [- \frac{\pi}{2},\frac{\pi}{2})$ because the density function is the same for $\bm{c}$ and $-\bm{c}$.

For each single circular (angular) observation, $\theta_k$, $k=1, \ldots, n$, we construct the complex vector of trigonometric moments defined as
$\bm{\hat{e}}^*_k \propto \left(1,e^{-i\theta_k},e^{-2i\theta_k}, \ldots, e^{-Mi\theta_k}\right)^\top$, where each vec\-tor is normalized to have unit norm.
After normalization and considering the vector of real and imaginary parts, $\bm{\hat{e}}$, these real vectors are in the parameter space of the NNTS densities, that is, the unit hypersphere of dimension $2M$ $S(2M)$ because the complex hypersphere $S_{\mathbb{C}}(M)$ is isomorphic with the real $2M+1$ unit hypersphere.

To transform the original dependent circular variables into linear variables, we consider the distance of the $\bm{\hat{e}}_k$ vectors to a geodesic with
known $\bm{a}$ and $\bm{d}$ vectors. The cosine between $\bm{\hat{e}}_{k}$ and the geodesic is given as follows:
\begin{equation}
	\begin{split}
\bm{\hat{e}}_{k}^\top\bm{\epsilon}_k & = \bm{\hat{e}}_{k}^\top\left\{ \bm{a}\cos(\phi_k) + \bm{d}\sin(\phi_k)\right\} \\
&= (\bm{\hat{e}}_{k}^\top\bm{a})\cos(\phi_k) + (\bm{\hat{e}}_{k}^\top\bm{d})\sin(\phi_k)
\end{split}
\end{equation}

which is maximized for
\begin{equation}
\hat{\phi}_k = \arctan \left(\frac{\bm{\hat{e}}_{k}^\top\bm{d}}{\bm{\hat{e}}_{k}^\top\bm{a}}  \right)
\end{equation}
for $k=1,2, \ldots, n$. Note that maximizing the cosine of the angle between two vectors is equivalent to minimizing the angle between those vectors. The nearest vector on the geodesic to $\bm{\hat{e}}_k$, denoted by $\bm{\epsilon}_k$, is given by
\begin{equation}
\bm{\epsilon}_k=
\frac{\bm{\hat{e}}_{k}^\top\bm{a}}{\sqrt{(\bm{\hat{e}}_{k}^\top\bm{a})^2+(\bm{\hat{e}}_{k}^\top\bm{d})^2}} \bm{a} +
\frac{\bm{\hat{e}}_{k}^\top\bm{d}}{\sqrt{(\bm{\hat{e}}_{k}^\top\bm{a})^2+(\bm{\hat{e}}_{k}^\top\bm{d})^2}} \bm{d} = w_{\bm{a}}\bm{a}+w_{\bm{d}}\bm{d}.
\end{equation}

Then, $\bm{\epsilon}_k$ is a weighted sum of $\bm{a}$ and $\bm{d}$ wherein the weight for vector $\bm{a}$, $w_{\bm{a}}$, increases with the cosine of the angle
between $\bm{\hat{e}}_{k}$ and $\bm{a}$, ($\bm{\hat{e}}_{k}^\top\bm{a}$). Here, $w_{\bm{a}}^2 + w_{\bm{d}}^2=1$, as implied by equation (\ref{geodesic}).

To complete the transformation of the circular variables into linear variables, it is necessary to obtain the values of $\bm{a}$ and $\bm{d}$. We consider the values of  $\bm{a}$ and $\bm{d}$ that maximize the sum of squared cosines of the angles between the vector of parameter estimates and the corresponding vector on the geodesic $SSC(\bm{a},\bm{d})$ defined as follows:

\begin{equation}
	\begin{split}
SSC(\bm{a},\bm{d}) & =
\sum_{k=1}^n \left(\bm{\hat{e}}_{k}^\top\bm{\epsilon}_k \right)^2
 =
\sum_{k=1}^n \left\{(\bm{\hat{e}}_{k}^\top\bm{a})^2 + (\bm{\hat{e}}_{k}^\top\bm{d})^2 \right\} \\
&=
\bm{a}^\top \left(\sum_{k=1}^n \bm{\hat{e}}_k\bm{\hat{e}}_{k}^\top\right)\bm{a} + \bm{d}^\top\left(\sum_{k=1}^n \bm{\hat{e}}_k\bm{\hat{e}}_{k}^\top\right)\bm{d} \\
& =  \bm{a}^\top\hat{E}\bm{a} + \bm{d}^\top\hat{E}\bm{d}
	\end{split}
\end{equation}
subject to constraints $\bm{a}^\top\bm{a} = 1$, $\bm{d}^\top\bm{d} = 1$, and $\bm{a}^\top\bm{d} = 0$ de\-fined in Equation (\ref{geodesic}). Here the estimates of $\bm{a}$ and $\bm{d}$,
$\bm{\hat{a}}$ and $\bm{\hat{d}}$,
correspond to the first and second unit norm eigenvectors of the $2M+1$ by $2M+1$ matrix $\hat{\mathit{E}}=\sum_{k=1}^n \bm{\hat{e}}_k\bm{\hat{e}}_{k}^\top$. In other words, we find
the great circle that best represents the points in the parameter hypersphere that correspond to the $n$ vectors containing the first $M$ trigonometric moments, $\bm{\hat{e}}_1, \ldots, \bm{\hat{e}}_n$. This is equivalent to maximizing the exponential of an orthogonal series density estimator (the NNTS model)
as per Clutton-Brock (1990). Moreover, to maximize $SSC$ with respect to the vectors defining the great circle, $\bm{a}$ and $\bm{d}$ and the
angles $\phi_k$ to obtain the definition of the linear variable $Y_k$ is consistent with the definition of the NNTS density in Equation (\ref{NNTS1}) as a squared
norm of the product of two complex vectors.

Finally, by maximizing $SSC(\hat{\bm{a}},\hat{\bm{d}})$ with respect to the rotation angle $\phi_k$, a new linear variable $Y_k$ is defined as follows:
\begin{equation}
Y_k = \tan\left(\arctan\left(\frac{\bm{\hat{e}}_{k}^\top\hat{\bm{d}}}{\bm{\hat{e}}_{k}^\top\hat{\bm{a}}}\right)\right) = \frac{\bm{\hat{e}}_{k}^\top\hat{\bm{d}}}{\bm{\hat{e}}_{k}^\top\hat{\bm{a}}}
\end{equation}
for $k = 1, 2, \ldots, n$. For example, in a linear regression case, the estimator of vector $\bm{\beta}=(\beta_1, \ldots, \beta_p)^\top$, $\hat{\bm{\beta}}$, is obtained as
$\bm{\hat{\beta}}=(X^{\top} X)^{-1}X^{\top}\bm{Y}$ where $\bm{Y}=(Y_1, Y_2, \ldots, Y_n)^\top$, and $\mathit{X}$ is the matrix of explanatory variables. The regression model for $\bm{Y}$ is expressed as follows:
\begin{equation}
\bm{Y} = X\bm{\beta} + \bm{e}
\end{equation}
where $\bm{e}$ is a vector of random errors. Here, an intercept term ($\beta_0$) was not included in the model because of the unknown link function (see Li and Duan, 1989).
Once the vector of estimates $\bm{\hat{\beta}}$ is obtained using linear regression, the prediction equation for the $\bm{c}$ parameter vector as a function of the given vector of covariates $\bm{x}$
is given as follows:
\begin{equation}
\bm{\hat{c}} = \bm{\hat{a}}\cos(\hat{\phi}) + \bm{\hat{d}}\sin(\hat{\phi})
\label{predictc}
\end{equation}
with the estimated rotation angle on the geodesic obtained as:
\begin{equation}
\hat{\phi} = \arctan(\bm{x}^\top\bm{\hat{\beta}}).
\end{equation}
In this sense, $\hat{\phi}$ corresponds to the inverse of the median of the fitted linear model. Thus, a density forecast is obtained that corresponds to an NNTS density with vector of pa\-ram\-e\-ters $\bm{\hat{c}}=(Re(\bm{c^*}),Im(\bm{c^*}))^\top$ given in Equation (\ref{predictc}). If a point prediction of the angle is required, the angle defined by the estimated value of the argument of the complex number $e^{i\theta}$
under the NNTS density forecast defined as
\begin{equation}
\hat{E}(e^{i\theta}) = \hat{E}\left\{\cos(\theta) + i\sin(\theta)\right\} = 2\pi\sum_{k=0}^{M-1}\hat{c}^*_{k}\bar{\hat{c}}^*_{k+1}
\end{equation}
can be used.

\subsection{Small-circle Regression on Unit Hypersphere}

A small circle in the unit 2$M$ hypersphere $S(2M)$ is defined as follows:
\begin{equation}
\bm{\epsilon} = \cos(\alpha)\bm{b} + \sin(\alpha)\left[\bm{a}\cos(\phi) + \bm{d}\sin(\phi)\right]
\end{equation}
with $\bm{b}$, $\bm{a}$, and $\bm{d}$ orthogonal unit vectors.
Similar to the great circle case, given the vectors of the first $M$ trigonometric moments corresponding to each of the circular observations, $\theta_1, \ldots, \theta_n$, represented as $\bm{\hat{e}}_1, \ldots, \bm{\hat{e}}_n$, the vector on the small circle that is nearest to the vector $\bm{\hat{e}}_k$ can be obtained as follows:
\begin{equation}
	\begin{split}
\bm{\epsilon}_k & =  \left\{
\frac{\bm{\hat{e}}_{k}^\top\bm{b}}{\sqrt{(\bm{\hat{e}}_{k}^\top\bm{b})^2+(\bm{\hat{e}}_{k}^\top\bm{a})^2+(\bm{\hat{e}}_{k}^\top\bm{d})^2}}
\right\}
\bm{b}  \\
& + \left\{
\frac{\sqrt{(\bm{\hat{e}}_{k}^\top\bm{a})^2+(\bm{\hat{e}}_{k}^\top\bm{d})^2}}{\sqrt{(\bm{\hat{e}}_{k}^\top\bm{b})^2+(\bm{\hat{e}}_{k}^\top\bm{a})^2+(\bm{\hat{e}}_{k}^\top\bm{d})^2}}
\right\}\\
& \times
\left\{
\frac{\bm{\hat{e}}_{k}^\top\bm{a}}{\sqrt{(\bm{\hat{e}}_{k}^\top\bm{a})^2+(\bm{\hat{e}}_{k}^\top\bm{d})^2}}\bm{a} +
\frac{\bm{\hat{e}}_{k}^\top\bm{d}}{\sqrt{(\bm{\hat{e}}_{k}^\top\bm{a})^2+(\bm{\hat{e}}_{k}^\top\bm{d})^2}}\bm{d}
\right\}
	\end{split}
\end{equation}

Similar to the great circle case, $\bm{\epsilon}_k$ is a weighted sum of $\bm{b}$, $\bm{a}$ and $\bm{d}$, $\bm{\epsilon}_k=w_{\bm{b}}\bm{b}+w_{\bm{a}}\bm{a}+w_{\bm{d}}\bm{d}$ with $w_{\bm{b}}^2+w_{\bm{a}}^2+w_{\bm{d}}^2=1$. Further, each weight increases with the cosine
of the vector $\bm{\hat{e}}_{k}$ with the corresponding vector defining the small circle.

The estimates of the vectors $\bm{b}$, $\bm{a}$, and $\bm{d}$ are obtained by maximizing the sum of squared cosines
$SSC(\bm{b}, \bm{a}, \bm{d})$ defined as follows:
\begin{equation}
	\begin{split}
SSC(\bm{b}, \bm{a}, \bm{d}) & =   \sum_{k=1}^n (\bm{\hat{e}}_{k}^\top\bm {\epsilon}_k)^2 =
\sum_{k=1}^n \left\{(\bm{\hat{e}}_{k}^\top\bm{b})^2 + (\bm{\hat{e}}_{k}^\top\bm{a})^2 + (\bm{\hat{e}}_{k}^\top\bm{d})^2  \right\}  \\
& =  \bm{b}^\top\hat{E}\bm{b} + \bm{a}^\top\hat{E}\bm{a} + \bm{d}^\top\hat{E}\bm{d}
	\end{split}
\end{equation}
The estimators of $\bm{b}$, $\bm{a}$, and $\bm{d}$ correspond to the first,
second, and third unit-norm eigenvectors of $\hat{E}=\sum_{k=1}^n \bm{\hat{e}}_k\bm{\hat{e}}_{k}^\top$.

The estimator of the angle $\alpha$, $\hat{\alpha}$, is obtained by max\-i\-miz\-ing the sum of the squared cosines, $SSC(\hat{\bm{b}}, \hat{\bm{a}}, \hat{\bm{d}})$ and is equal to:
\begin{equation}
\hat{\alpha}=\frac{1}{2}\arctan \left(\frac{2\sum_{k=1}^n \sign(\bm{\hat{e}}_{k}^\top\hat{\bm{b}})
\sqrt{(\bm{\hat{e}}_{k}^\top\bm{a})^2+(\bm{\hat{e}}_{k}^\top\bm{d})^2}}{
 \sum_{k=1}^n \left\{ (\bm{\hat{e}}_{k}^\top\bm{b})^2-(\bm{\hat{e}}_{k}^\top\bm{a})^2-(\bm{\hat{e}}_{k}^\top\bm{d})^2 \right\}}\right).
\end{equation}
Analogous to the great-circle case, the new linear dependent variable $Y_k$ corresponding to $\theta_k$ is obtained by maximizing $SSC(\hat{\bm{b}}, \hat{\bm{a}},\hat{\bm{d}})$ with respect to angle $\phi_k$ as
\begin{equation}
Y_k = \tan\left(\arctan\left(\frac{\bm{\hat{e}}_{k}^\top\hat{\bm{d}}}{\bm{\hat{e}}_{k}^\top\hat{\bm{a}}}\right)\right) = \frac{\bm{\hat{e}}_{k}^\top\hat{\bm{d}}}{\bm{\hat{e}}_{k}^\top\hat{\bm{a}}}
\end{equation}
because a small circle is the translation of a great circle. Thus, to discriminate between fitted great and small circle models,
the relation between the estimated $\alpha$ and $\phi$ angles can be evaluated. The great circle is nested in a small circle. For example, for a $\phi$ angle equal to zero, the small circle transforms to a great circle defined by the vectors $\bm{b}_0$ and $\bm{c}_0$. Another possible situation where the small circle transforms to a great circle is when $\alpha=\frac{\pi}{2}$ or $\phi=\frac{\pi}{2}$.
For a particular vector of covariates $\bm{x}$, a density forecast is obtained using the NNTS density with the $\bm{\hat{c}}$ vector of parameters expressed as
\begin{equation}
\bm{\hat{c}} = \cos(\hat{\alpha})\bm{\hat{b}} + \sin(\hat{\alpha})\{\bm{\hat{a}}\cos(\hat{\phi}) + \bm{\hat{d}}\sin(\hat{\phi})\}
\label{smallcircleeqangle}
\end{equation}
For the estimation of the geodesic rotation angle, the fol\-low\-ing two cases must be considered (Presnell et al. 1998):
\begin{equation}
\hat{\phi} = \arctan(\bm{x}^\top\bm{\hat{\beta}}) \mbox{ if } \bm{\hat{e}}_{k}^\top\hat{\bm{a}} > 0
\end{equation}
and
\begin{equation}
\hat{\phi} = \left\{\arctan(\bm{x}^\top\bm{\hat{\beta}}) + \pi\right\} \bmod 2\pi \mbox{ if } \bm{\hat{e}}_{k}^\top\hat{\bm{a}} \le 0
\end{equation}
because, contrary to the great-circle case, $\phi$ and $\phi + \pi$ do not generate the same NNTS density.
Thus, in the small-circle case, two models are obtained. The first one for $\bm{\hat{e}}_{k}^\top\hat{\bm{a}}<0$ and the second one for $\bm{\hat{e}}_{k}^\top\hat{\bm{a}} \ge 0$ that produce different NNTS density forecasts. In case of needing a unique NNTS density forecast,
the resultant vector of the two $\bm{\hat{c}}$ parameter vectors for the cases $\bm{\hat{e}}_{k}^\top\hat{\bm{a}} > 0$ and $\bm{\hat{e}}_{k}^\top\hat{\bm{a}} \le 0$ can be used.

\subsection{Goodness of Fit and Probability Integral Transform Validation}

The goodness of fit of the proposed model comprises two elements. The first measures the goodness of fit of the projection of the original
circular dependent variables onto the pa\-ram\-e\-ter hypersphere in terms of the cosine distances of the real vectors with the first $M$ trigonometric moments
$\bm{\hat{e}}_1, \bm{\hat{e}}_2, \ldots, \bm{\hat{e}}_n$ to the great (small) circle. It is possible to define
this measure in terms of the sum of the squared cosines of the vectors to their nearest points on the great (small) circle
\begin{equation}
R^2_{cos} =\frac{SSC}{n}
\end{equation}
because the maximum value of SSC is equal to $n$ (i.e., the number of observations) when all trigonometric moments vectors are on the great (small) circle. Then, the separation
angles are equal to zero, and the cosine is equal to one. Note that $0 \le R^2_{cos} \le 1$ and a greater $R^2_{cos}$ value indicate that the great
(small) circle better represents vectors $\bm{\hat{e}}_1, \bm{\hat{e}}_2, \ldots, \bm{\hat{e}}_n$.

The second element is related to the goodness of fit of the regression
model fitted to the transformed variable $Y$. Here, for example, the coefficient of determination $R^2$ (squared correlation between observed and fitted $Y$ values) can be used when employing linear regression.

We focus on fitting NNTS densities; thus, validation of the models should be performed by considering techniques related to density forecasting. Here,
we use the probability integral transform (PIT) to validate the fitted models. Fol\-low\-ing Diebold et al. (1998), we consider uniformity tests for the
PITs of the fitted models evaluated at the observed values. To perform the PIT calculations, we must consider
that the model used to determine the (geodesic) rotation angle on the great (small) circle uses the transformed variable $Y$, which is defined as follows:
\[
Y=\frac{\bm{\hat{e}}^\top \bm{d}}{\bm{\hat{e}}^\top \bm{a}}
\]
and that the estimated value of the rotation angle, $\phi$, obtained as
\[
\hat{\phi}=\arctan(\bm{x}^\top\hat{\bm{\beta}}),
\]
is equivalent to considering the median of the inverse transformation from the linear fitted model. We also considered other inverse transformations, such as the inverse of the expected value considering a second-order Taylor series; however, the best results were obtained using the inverse of the median.

We proceed to find the $\bm{c}$ parameter vector that cor\-re\-sponds to the rotation angle using Equation (\ref{predictc}) (Equation (\ref{smallcircleeqangle})) defining the great (small) circle, and calculate the prob\-a\-bil\-i\-ty integral in the corresponding observed value.

By applying this procedure to the $n$ observed values, $\theta_1, \theta_2, \ldots, \theta_n$, we obtain the vector of PIT values that, if the
model is a good approximation of the process that generated the data, should be $i.i.d.$ observations from a circular uniform density function.

To evaluate the uniformity of the PIT values, the circular uniformity omnibus tests of Kuiper (1960), Watson (1961) and, the range test were used, as suggested by Mardia and Jupp (2000) to test the goodness of fit for circular data. The p-values of these tests were obtained using the $R$ $circular$ package (Agostinelli and Lund 2017).

When selecting the best value of $M$ in terms of the uniformity tests of the PIT values, the p-values of the range test can be adjusted for
multiple hypothesis testing by using the procedure of Benjamini and Hochberg (1995).

\section{Examples}

\subsection{Simulated Data}

A design matrix of explanatory variables with 1000 rows and five columns was simulated. The entries in the first column ($X_1$) corresponded to realizations of a binary variable centered at zero, with values of -1 and 1. The second column ($X_2$) contained realizations from a discrete variable with values ranging from 1 to 40. Finally, the last three columns ($X_3$, $X_4$, and $X_5$) included realizations from normally distributed random variables with variance equal to one and means of 4, 6, and 8, respectively. In addition, the last four variables ($X_2$, $X_3$, $X_4$, and $X_5$) were standardized to have a mean of zero and unit variance. Moreover, for the values of the beta regression parameters, four different vectors of beta parameters were considered. Considering the problem
of the unknown link function (Li and Duan, 1989), the beta vectors do not include the intercept term $\beta_0$.
The first vector is equal to $\bm{\beta}_1=(0,0,0,0,0)^\top$, where all beta regression parameters are equal to zero; the second to $\bm{\beta}_2=(0.3,0.2,0.15,0.2,0.3)^\top$; the third to $\bm{\beta}_3=(0,0,0,0,0.3)^\top$; and the fourth to $\bm{\beta}_4=(x,x,x,x,0.3)^\top$, where
$x$ implies that the variable associated with the corresponding position of the vector was not included in the simulated dataset; thus, for the $\bm{\beta}_4$ case,
only the column $X_5$ was included in the design matrix.

For each value $M= 1, 2, 3, \ldots, 8$ sample sizes, $n$, equal to 25, 50, 100, 200, 500, and 1000 were considered.
Further, for each possible combination of $M$ and $n$, the vectors defining the great (small) circle, $\bm{a}$, and $\bm{d}$
($\bm{b}$, $\bm{a}$, and $\bm{d}$) were simulated by simulating 5000 realizations from an NNTS density with $M$ components and a randomly selected $\bm{c}$ parameter. Consequently,  the matrix of trigonometric moments was constructed to obtain the first two (three) eigenvectors that corresponded to the simulated vectors defining the great (small) circle. Further, for a sample size of $n$, the first $n$ rows of the design matrix are used. In case of the $k$-th row of the simulated design matrix, in the great (small) circle case a realization from an NNTS density with $\bm{c}=\bm{a}\cos(\phi_k) + \bm{d}\sin(\phi_k)$
($\bm{c}=\cos(\alpha)\bm{b}+ \sin(\alpha)(\bm{a}\cos(\phi_k)+\bm{d}\sin(\phi_k))$) vector of parameters was simulated to obtain the vector of values of the circular dependent variable. This, coupled with the first $n$ rows of the simulated design matrix corresponded to the simulated dataset.

For each of the four vectors of beta parameters and possible combination of $M$ and $n$, 100 simulated datasets were generated. Subsequently, the proposed estimation methodology was applied to each simulated dataset to obtain the estimates of the beta regression parameters considering two cases. One wherein the simulated vectors defining the great (small) circle are not estimated and considered as known, and the second wherein these are estimated with the beta regression parameters. In Tables \ref{Table04simulationGC} and \ref{Table05imulationSC}, the results for $M=1,2, \ldots, 5$ are presented to estimate the beta parameters for the great and small circle models, respectively. The simulations indicated that for the estimated eigenvectors case, when $M=6, 7, \mbox{ or } 8$, biased results were produced because it was not always possible to obtain reliable estimates of the eigenvectors of the matrix of trigonometric moments owing to the high dimensionality of the matrix, and the estimates for $M=6,7 \mbox{ and } 8$ were not aggregated to the results. Thus, when the estimated eigenvectors are not good estimates of the real eigenvectors, the PIT uniformity tests reject the null hypothesis of uniformity, which is good for practical applications. In contrast, when the estimated eigenvectors are good estimates of the real eigenvectors, the PIT uniformity tests reject the null hypothesis of uniformity in the expected rejection rate that corresponds to the selected significance level.

For the case 1 wherein $\bm{\beta}_1=(0,0,0,0,0)^\top$ for the great circle regression, the simulations presented in Table \ref{Table04simulationGC} show that in the great circle case, the percentage of the 100 simulations where the null hypothesis of each parameter is equal to zero, the rejection rate (RR), approximates the selected 5\% significance level both in the known and estimated eigenvector cases. This implies that the proposed methodology can detect non-significant explanatory variables at the expected 5\% significance level. In addition, the acceptance rates (AR) for the circular uniformity tests of the PIT values are adequate for the Kuiper, Watson, and range tests when the eigenvectors are known. However, when the eigenvectors are estimated, only the range test presents adequate values for large sample sizes. Moreover, in case of the small sample sizes, the beta parameter estimates are larger than for small sample sizes, although the RRs are correct. The corresponding results for $\bm{\beta}_0=(0,0,0,0,0)^\top$ in Table \ref{Table05imulationSC} are similar to those for case 1 in Table \ref{Table04simulationGC}, although the ARs for the Kuiper and Watson circular uniformity tests in the estimated eigenvectors case are very low. This pattern is observed in all estimated eigenvector cases in Tables \ref{Table04simulationGC} and \ref{Table05imulationSC}; thus, using the range test is recommended.

The Case 2 with $\bm{\beta}_2=(0.3,0.2,0.15,0.2,0.3)^\top$ shows  the effect on the beta estimates of the unknown link function (Li and Duan, 1989). For example, in Table \ref{Table05imulationSC} with known eigenvectors and sample size equal to 1000, the beta estimates are proportional to the true values with a proportionality constant of approximately 2/3. In addition, as expected, the RRs were smaller for smaller beta parameters. Similar results were obtained for the corresponding case in the small circle regression as presented in Table \ref{Table05imulationSC}, with a smaller proportionality constant of approximately 3/4. For Case 2, the results for the estimated eigenvectors are similar in the great and small circle regressions.

For the cases 3 and 4, the results presented in Tables \ref{Table04simulationGC} and \ref{Table05imulationSC} confirm that the proposed methodology correctly identified the significant beta parameter in the presence of other non-significant beta parameters. For the known eigenvectors, the RRs increased for Case 4, wherein non-significant variables have been eliminated, when compared to Case 3. For the estimated eigenvectors, the RRs for Cases 3 and 4 are similar in both Tables \ref{Table04simulationGC} and \ref{Table05imulationSC}.

Other cases were simulated and the following recommendations were found: the proposed methodology is safe to use for values of M up to 5 for which the estimates of the eigenvectors of the matrix of trigonometric moments are reliable. For values of $M$ greater than 5, the PIT uniformity tests should be used to validate that the
estimates of the eigenvectors are reliable. The range test is recommended for the circular uniformity test of PIT values to validate the fit of the models
specially for large sample sizes.

\subsection{Blue Periwinkles' Direction and Distance Data}

\begin{figure}[ht]
\includegraphics[scale=.5, bb= 0 0 504 504]{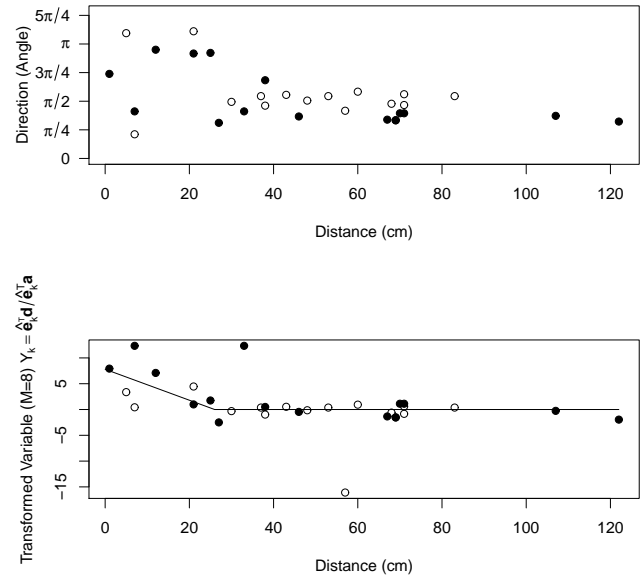}
\centering
\caption{\label{graph01datablueperiwinkles} Blue periwinkle data. The top plot shows the original data on the angle taken and distance travelled by 31 blue periwinkles one day after transplantation (15 individuals) and four days after transplantation (16 individuals). The bottom plot shows the scatterplot of the transformed variable when considering a small-circle nonnegative trigonometric sums (NNTS) model with $M=8$. The filled points correspond to the case where $\bm{\hat{e}}_{k}^\top\hat{\bm{a}} \le 0$, and non-filled points correspond to the case where $\bm{\hat{e}}_{k}^\top\hat{\bm{a}}>0$.}

\end{figure}

Fisher and Lee (1992) and Fisher (1993) analyzed the com\-pass direction ($\theta$) (relative to north) and distance ($x$) travelled by 31 periwinkles (small sea snails \emph{Nodilittorina unifasciata}) after transplanting downshore from their normal living height. The sea was at an angle of approximately 275$^\circ$.
A total of 15 observations corresponded to periwinkles mea\-sured one day after transplantation,
and the remaining 16 observations were measured four days later. Note that no significant differences were observed between the behaviors of these two groups. The angles of the compass directions are plotted as a function
of the traveled distance in Figure \ref{graph01datablueperiwinkles}.

For the directions taken by the blue periwinkles, Fisher (1992, 1993) considered two von Mises models. The first model considers the mean parameter as a
function of the distance traveled. The second (mixed) model considers the mean and concentration parameters as functions of the dis\-tance traveled. Similar to our methodology, the distance was included in the mean parameter using the tangent function of the linear predictor. In addition, an exponential function of the linear predictor was used for the concentration parameter. With the von Mises model,
Fisher (1993) concluded that clear evidence exists to support that both the mean and con\-cen\-tra\-tion parameters of the von Mises model for the di\-rec\-tions depend on distance.
For the mean parameter case, the estimated coefficient for the distance was negative, with a global mean of 2.044 (97$^\circ$). In the concentration parameter case,
the estimated beta parameter associated with the dis\-tance was positive, which implies that concentration in\-creased with distance.

Presnell et al. (1998) analyzed blue periwinkle data using a spherically projected multivariate linear (SPML) model that, for the circular case, corresponds
to using an angular normal distribution for the dependent variable. Presnell et al. (1998) compared the fit of their model to that of Fisher and Lee (1992), and concluded that the Fisher and Lee model suffered from serious problems related to numerical op\-ti\-miza\-tion of the likelihood function and
parameter non-iden\-ti\-fi\-a\-bil\-i\-ty. Thus, they considered that the values of the max\-i\-mized loglikelihoods should not be compared.

Recently, in the context of modeling directions taken by animals, Rivest et al. (2016) considered circular regression models for dependent and explanatory
circular variables, where the conditional mean was modeled using the inverse tangent
function and von Mises distributed errors with the option that the variance of the errors can be a function of the length of the conditional mean vector.

Figure \ref{graph01datablueperiwinkles} shows a dispersion plot of the original directions and distances. As can be seen, the angle (direction) tended to zero with distance. Table \ref{Table01blueperiwinkles} presents the estimates of the regression coefficients for the geodesic angle of rotation, the average squared cosine distances to the great (small) circle $R^2_{cos}$, the p-values of the range, Kuiper and Watson tests
of circular uniformity for the probability integral transforms and the loglikelihood values for the considered models with different values for parameter $M$ ($M=1, 2, \ldots, 8$), that is, the number of terms in the sum defining the NNTS model in Equation \ref{NNTS1}. In addition, the parameter estimates, the p-values of the circular uni\-for\-mi\-ty
tests, and the loglikelihood of the Fisher and Lee (1992) mean and mixed models, the Presnell et al. (1998) SPM mixed model, and Rivest et al. (2016) full and re\-cip\-ro\-cal mixed models are
also given in Table \ref{Table01blueperiwinkles}.

The $M=0$ case, which corresponds to a uniform circular density (see Table \ref{Table01blueperiwinkles}), shows the smallest value of the log\-like\-li\-hood, and the rnage, Kuiper and Watson tests clearly reject the circular uniformity of the
PIT values, which indicates that the directions taken by the blue periwinkles were not uniformly distributed on the circle.

After examining the dispersion plot of the trans\-formed linear variable $Y$ versus the distance in the right panel of Figure \ref{graph01datablueperiwinkles}, the regression $Y=\beta_1I(distance \le 27)(distance-27) + error$ was proposed because the values after $distance=27$ look around the horizontal
axis. Here, the indicator function $I(distance \le 27)$ is equal to one if $distance \le 27$ and zero if $distance > 27$. The small circle fitted model with $M=8$ is shown in the bottom panel of Figure \ref{graph01datablueperiwinkles}. When considering this regression equation, the small circle model with $M=8$ had a loglikelihood of -19.786, and the p-values of the range, Kuiper and Watson tests were greater than 10\%, which means that the null hypothesis of circular uni\-for\-mi\-ty of the PIT values was not rejected and the goodness of fit of the model was confirmed. The small circle model with $M=8$ has the largest p-value for the range test even after adjusting the p-values for multiple hypothesis testing by using the procedure of Benjamini and Hochberg (1995). For the small circle model with $M=5$ and regression equation $Y=\beta_1I(distance \le 27)(distance-27) + error$, the p-values of the
Kuiper and Watson tests were less than 0.01 although the range test yielded a p-value greater than 0.05 but less than 0.10 even after adjusting the p-values for multiple hypothesis testing. Thus, the goodness of fit of this model was considered inferior to the one with $M=8$, although it demonstrated the largest loglikelihood value equal to -16.902 and the largest $R^2$ value followed by the model with $M=8$.

\begin{figure}[ht]
	
\centering
\includegraphics[scale=0.3, bb= 0 0 504 504]{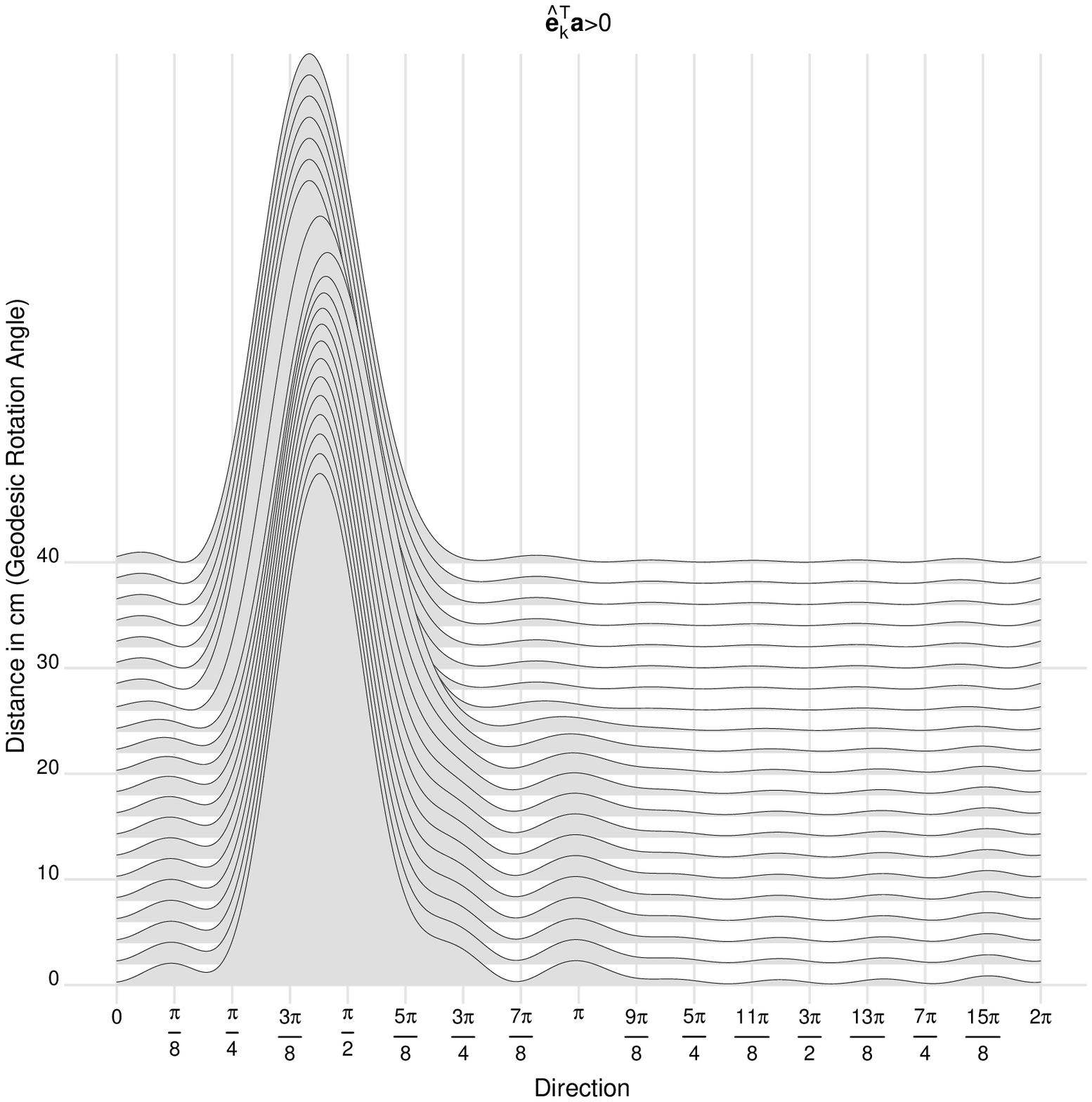}
\includegraphics[scale=0.3, bb= 0 0 504 504]{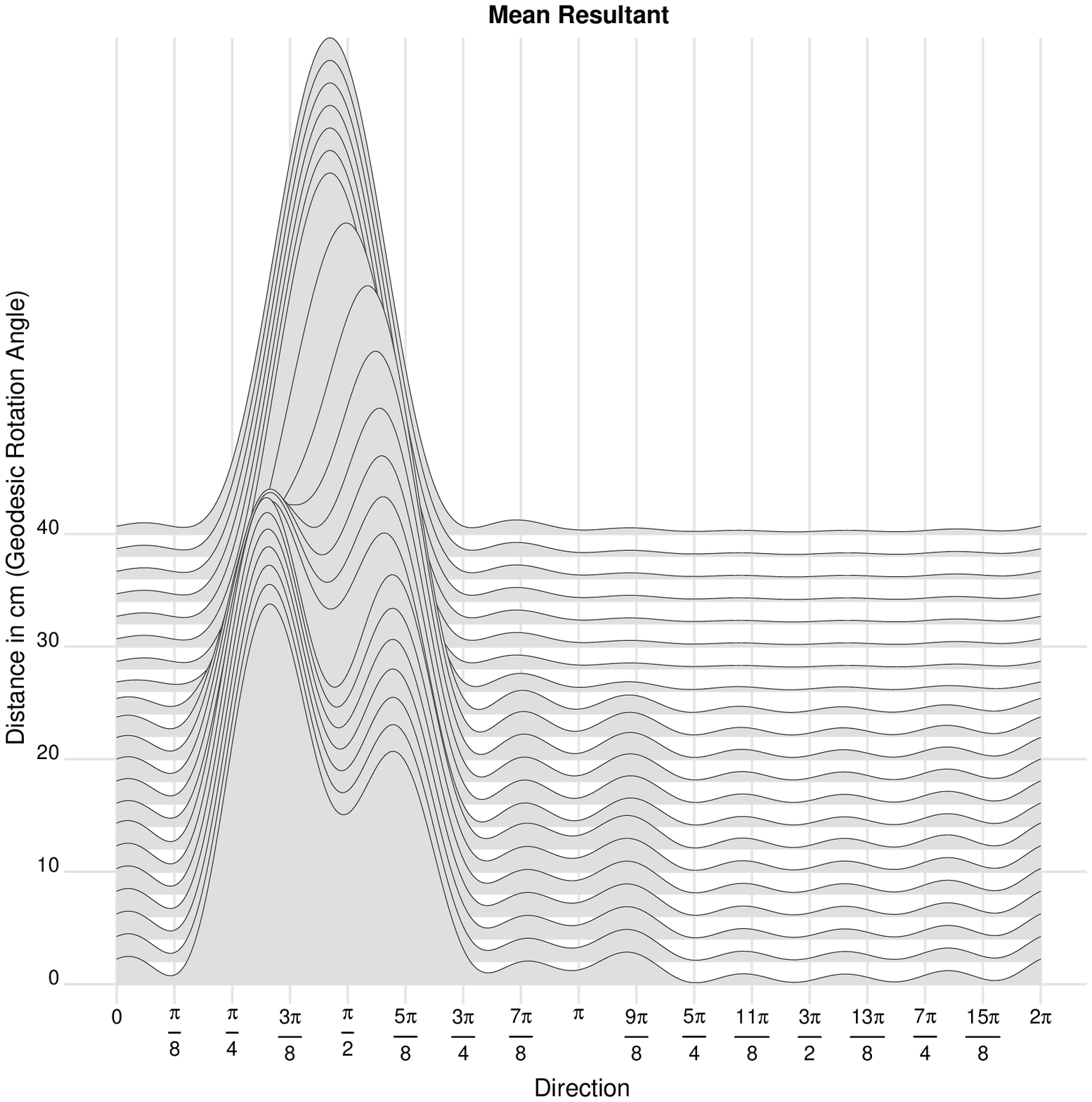}

\includegraphics[scale=0.3, bb= 0 0 504 504]{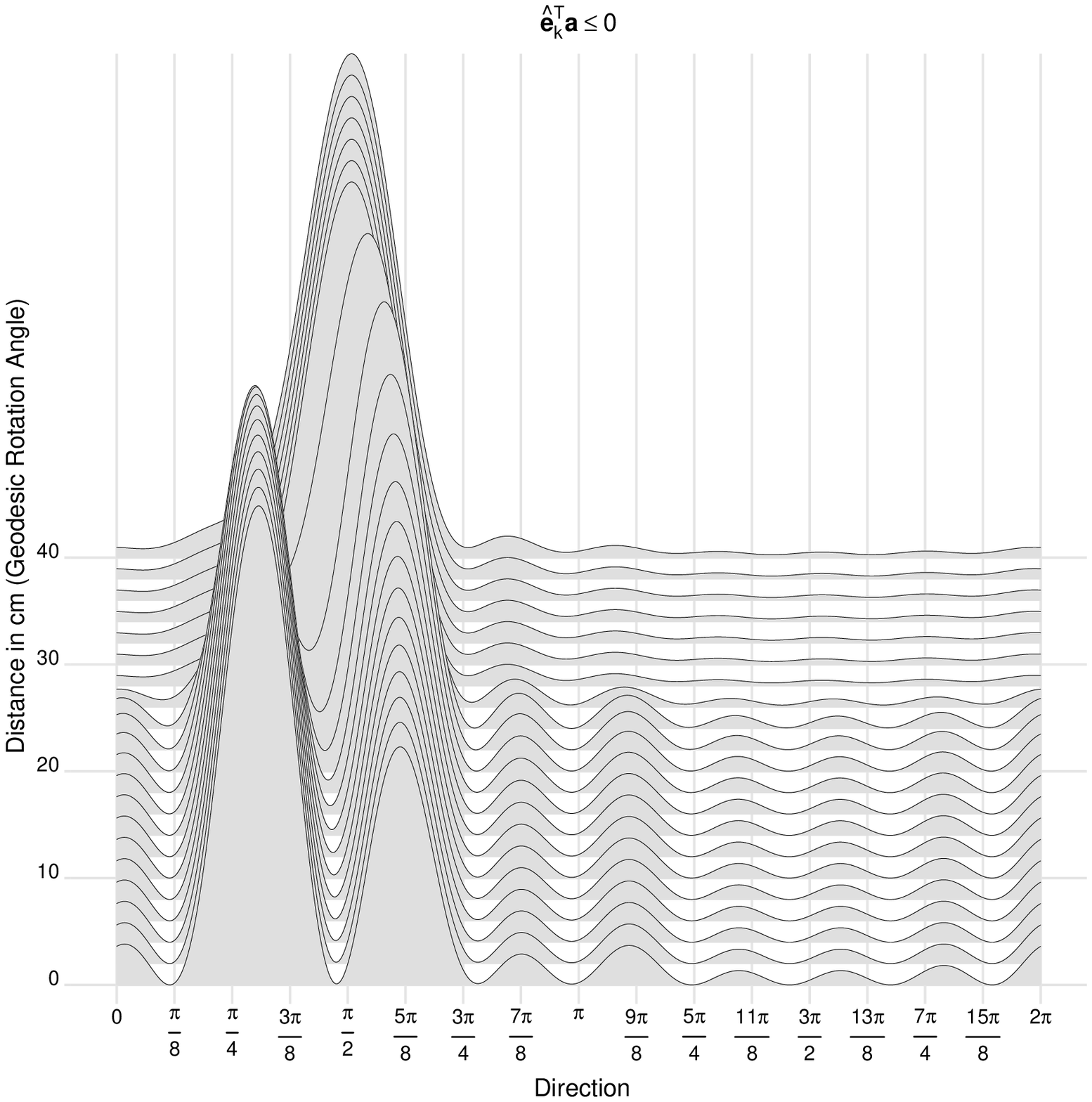}
\includegraphics[scale=0.3, bb= 0 0 504 504]{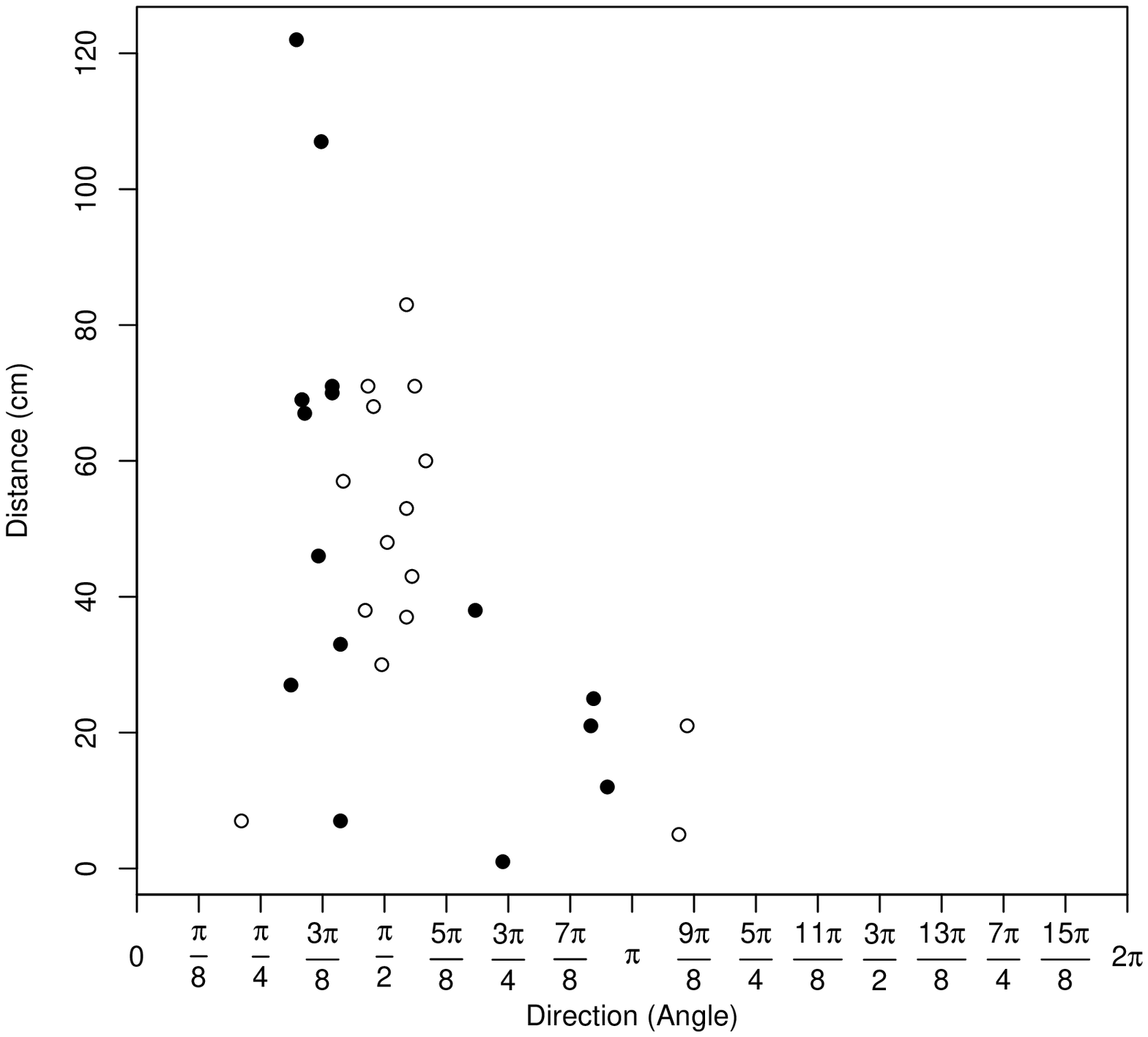}

\caption{\label{graph01datablueperiwinklescosine} Blue periwinkle data. Density forecasts for small-circle nonnegative trigonometric sums (NNTS) model with $M=8$ are shown. The top left plot corresponds to
	$\bm{\hat{e}}_{k}^\top\hat{\bm{a}} > 0$, and the bottom left plot corresponds to $\bm{\hat{e}}_{k}^\top\hat{\bm{a}} \le 0$. The top right plot
	corresponds to the case of the mean resultant length of the $\bm{\hat{e}}_{k}^\top\hat{\bm{a}}>0$ and $\bm{\hat{e}}_{k}^\top\hat{\bm{a}} \le 0$ cases.
	The bottom right plot shows the scatterplot of the original data with direction in the horizontal axis.
	Here, filled points correspond to $\bm{\hat{e}}_{k}^\top\hat{\bm{a}} \le 0$ and non-filled points correspond to $\bm{\hat{e}}_{k}^\top\hat{\bm{a}}>0$.}

\end{figure}

Figure \ref{graph01datablueperiwinklescosine} shows the predicted NNTS density functions for the best small-circle model ($M=8$ and $\hat{Y}= -0.300I(distance\le 27)(distance-27)$) for cases $\bm{\hat{e}}_{k}^\top\hat{\bm{a}}>0$ (top left plot) and
$\bm{\hat{e}}_{k}^\top\hat{\bm{a}} \le 0$ (bottom left plot), as well as the resultant length from the previous two cases (top right plot). In addition, the
bottom right plot of Figure \ref{graph01datablueperiwinklescosine} shows the dispersion plot of the original data with the direction on the horizontal axis and
distance on the vertical axis, to visually confirm the fit of the forecasted NNTS densities in the other three plots.

The NNTS density forecasts in the top left plot of Figure \ref{graph01datablueperiwinklescosine} for distances less than or equal to 27 are multimodal, with three main modes. For distances greater than 27, the NNTS density forecast was approximately unimodal with a mode at approximately 1.31 (75$^\circ$). For the NNTS density forecasts in the bottom left plot of Figure \ref{graph01datablueperiwinklescosine} satisfying a distance of less than or equal to 27, there are eight modes but two main modes. For distances greater than 27, the main mode was approximately $\frac{\pi}{2}$. Finally, the resultant mean NNTS density forecasts in the top right plot of Figure
\ref{graph01datablueperiwinklescosine} are presented, wherein for distances less than or equal to 27, the densities are multimodal with a high concentration
between $\frac{\pi}{8}$ and $\frac{3\pi}{4}$ and low concentrations cen\-tered at $\frac{7\pi}{8}$ and $\frac{9\pi}{8}$. For distances greater than 27, the fore\-cast\-ing
density exhibits a main mode below $\frac{\pi}{2}$. The resultant mean NNTS density forecast shows a good fit to most points in the scatterplot of the
bottom right plot of Figure \ref{graph01datablueperiwinklescosine}. Sim\-i\-lar\-ly, the density forecasts in the top (bottom) left plot of Figure \ref{graph01datablueperiwinklescosine} show a good fit to most non-filled (filled) points in the scatterplot in the bottom right plot.

\begin{figure}[ht]

\centering
\includegraphics[scale=0.65, bb= 0 0 504 504]{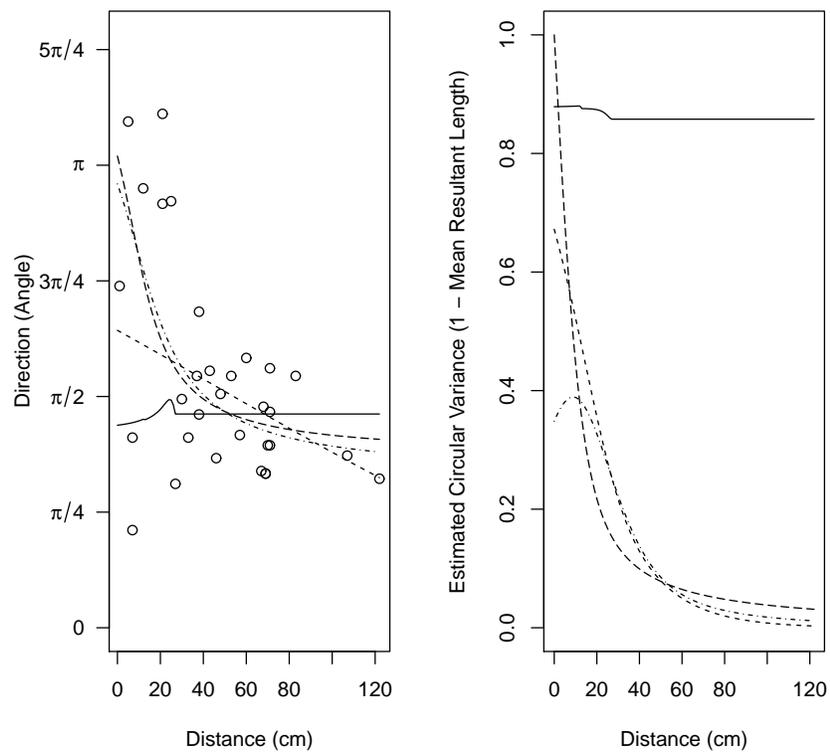}

\caption{\label{graph01datablueperiwinklesMEANFUNCTIONS} Blue periwinkle data. Estimated mean and circular variance functions for great-circle nonnegative trigonometric sums (NNTS) resultant mean model with $M=8$ (solid line), Fisher and Lee (1992) mixed model (dashed line), Presnell et al. (1998) SPM mixed model (dot-dash line) and Rivest et al. (2015) mixed model (long-dash line) are shown.}
\end{figure}

Figure \ref{graph01datablueperiwinklesMEANFUNCTIONS} shows the mean (argument of the first trigono\-met\-ric moment) and circular variance (one minus the mean resultant length, which is the
modulus of the first trigono\-met\-ric moment) functions as functions of the distance for the NNTS model,
the Fisher and Lee (1992) mixed model, the Presnell et al. (1998) mixed model, and the Rivest et al. (2015) full mixed model. In terms of the mean and circular variance functions,
the NNTS model has a resultant mean function of approximately 1.43 (82$^\circ$), which increases from a distance equal to zero to a distance equal to 27 and stays
constant at a distance of 27. Essentially, on average, the blue periwinkles moved in an approximate direction that corresponds to the negative of the angle the sea was located (275$^\circ$) and, for
distances less than 27, the density forecasts show many different modes with two main modes, which con\-verge toward a nearly unimodal density forecast for distances greater than 27.
The circular variance function of the NNTS mean resultant model demonstrates an almost constant pat\-tern with values of approximately 0.86. In contrast to the NNTS model, the other models present mean and circular variance functions that decrease with distance. As indicated by Rivest et al. (2015), this is an artifact of the joint
modeling of the mean and variance in their mixed model; however, it also applies to the mixed models of Fisher and Lee (1992) and Presnell et al. (1998).
Thus, the NNTS model presents an alternative explanation wherein the circular variance does not change with distance.

\subsection{Time Series of Wind Directions}

Fisher (1993) (see Cameron 1983) and Fisher and Lee (1994) analyzed the time series of 72 hourly wind directions over four days at a location on Black Mountain, Australian Capital Territory, to calibrate three anemometers.
They suggested two strategies to analyze circular time series. For noisy series, models should be fitted to the circular data directly, and
for less noisy series, the circular data should be transformed to linear data; thus, common linear time series methods should be applied. Note that the second
strategy is similar to the one we are proposing in this paper. By considering the first strategy, a circular autoregressive model of first order CAR(1)
was considered to fit the data. The CAR(1) model is con\-struct\-ed by considering that the
conditional distribution of the circular variable at time $t$ conditional on its previous values at times, $t-1$, $t-2$, $\ldots$ follows a von Mises
dis\-tri\-bu\-tion with mean direction $\mu_t$ and fixed concentration pa\-ram\-e\-ter $\kappa$. The conditional mean direction $\mu_t$  is constructed by con\-sid\-er\-ing
a link function (e.g., the inverse tangent func\-tion). Here, the estimated value of $\hat{\kappa}$ exceeds two, and the second strategy is applied by transforming
the circular data into linear data using a probit link, which implies an au\-tore\-gres\-sive model
of order one AR(1) for the transformed data. As stated by Fisher and Lee (1994), the LAR(1) model has unknown density for the original circular data, and the loglikelihood and PIT values cannot be calculated. Thus, we do not consider the LAR(1) model.
The estimated mean directions obtained by the CAR(1) and LAR(1) models in the aforementioned study were 5.05 (289.5$^{\circ}$)
and 5.19 (297.2$^{\circ}$), respectively.

\begin{figure}[ht]

\centering
\includegraphics[scale=.65, bb= 0 0 504 504]{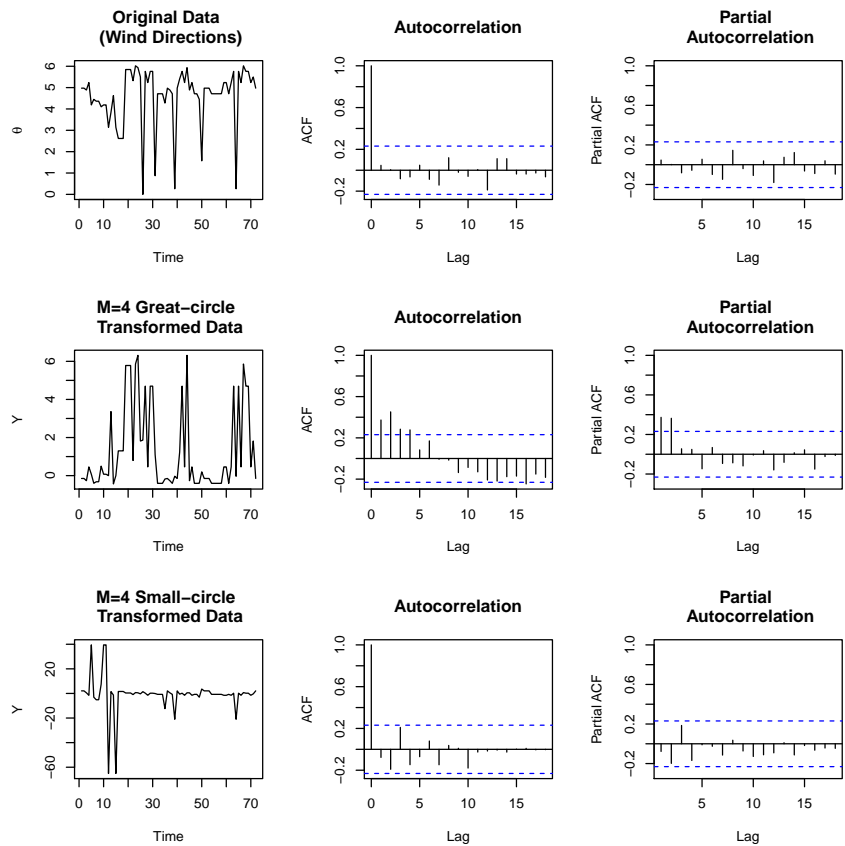}

\caption{\label{graph02acfpacf} Wind direction data. The original and transformed data for the great- and small-circle nonnegative trigonometric sums (NNTS) models with $M=4$ are shown including autocorrelation (ACF) and partial
	autocorrelation (PACF) functions.}

\end{figure}

To compare the proposed method to the CAR(1) model results obtained by Fisher (1993) and Fisher and Lee (1994), we fit the NNTS models to 72
observations. Figure \ref{graph02acfpacf} shows the time-series plots of the original data, along with the great-circle and small-circle transformed data
as their sample au\-to\-cor\-re\-la\-tion (ACF) and partial autocorrelation (PACF) plots when considering great-circle and small-circle NNTS mod\-els with $M=4$. For the great-circle transformed data, the ACF and PACF are consistent with the AR(2) model, and for the small-circle transformed data, they are consistent with a white noise process.

Table \ref{Table02winddirections1to72AR1} includes the NNTS models by fitting AR(1) mod\-els with zero mean
and NNTS densities with $M=1, 2, \ldots, 5$. As can be seen, among the NNTS great-circle AR(1) models, the model with $M=4$ shows the greatest loglikelihood value (-78.050) and the largest p-value for the range test (0.183). In addition, this model confirms that the PIT cir\-cu\-lar uniformity range, Kuiper and Watson tests do not reject the null hypothesis of circular uniformity. This model could be a candidate for our final model, and we note that its log\-like\-li\-hood is greater than that of the CAR(1) model (-81.370). The autoregressive coefficient is significant with the estimate equal to 0.5276, which is similar to the value reported by Fisher and Lee (1994). For all NNTS small-circle models, the range, Kuiper and Watson tests of circular uni\-for\-mi\-ty rejected the null hypothesis of circular uni\-for\-mi\-ty for the PIT values at a 10\% significance level.
For the CAR(1) model, the range test did not reject the null hypothesis of uniformity contrary to the Kuiper and Watson tests.

By inspecting the ACF and PACF plots for the residuals of the NNTS great-circle AR(1) model with $M=4$, we found that it is necessary to include a second autoregressive term, as observed in Figure \ref{graph02acfpacf}. The last row
of Table \ref{Table02winddirections1to72AR1} includes the results for the NNTS great-circle AR(2) model with $M=4$, which obtained a loglikelihood of -73.967 (the
greatest loglikelihood value of all considered models), and the range, Kuiper and Watson tests do not reject the null hypothesis of circular uniformity of the PIT values at a 10\% significance level. In addition, the ACF and PACF of the residuals present the pattern of a white noise process, which validates this model.

\begin{figure}[ht]
	
\centering
\includegraphics[scale=.3, bb= 0 0 504 504]{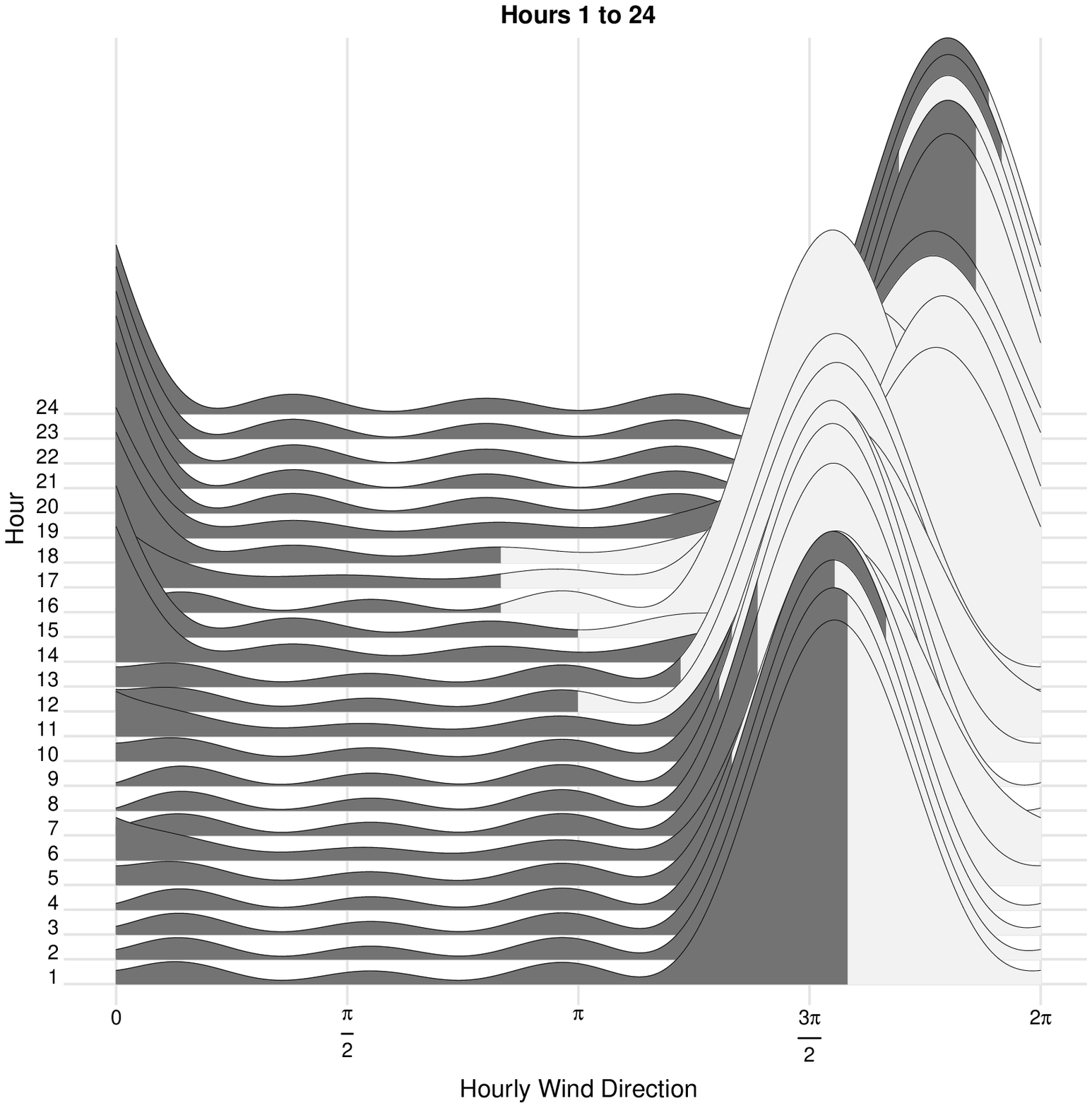}
\includegraphics[scale=.3, bb= 0 0 504 504]{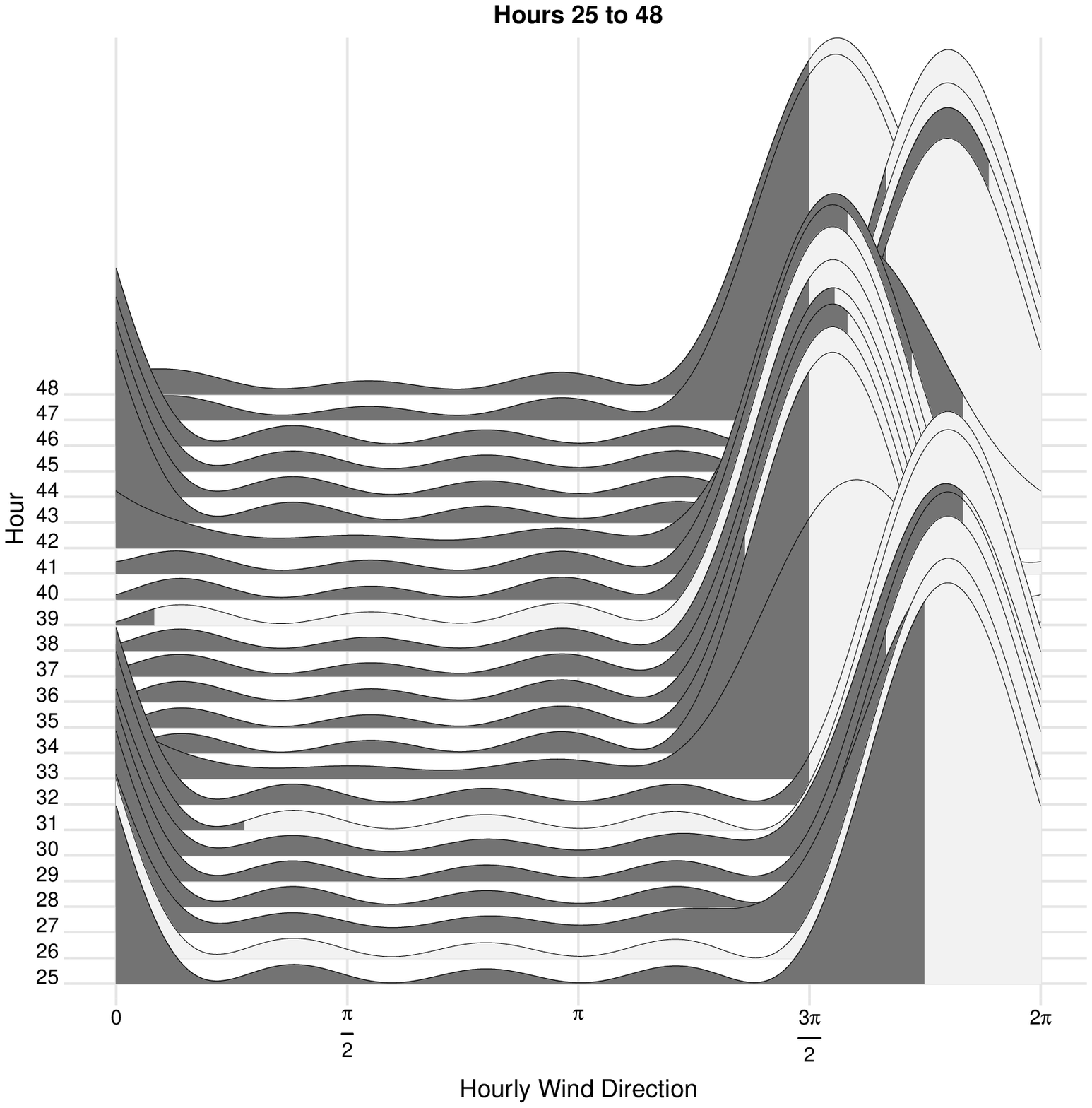}

\includegraphics[scale=.3, bb= 0 0 504 504]{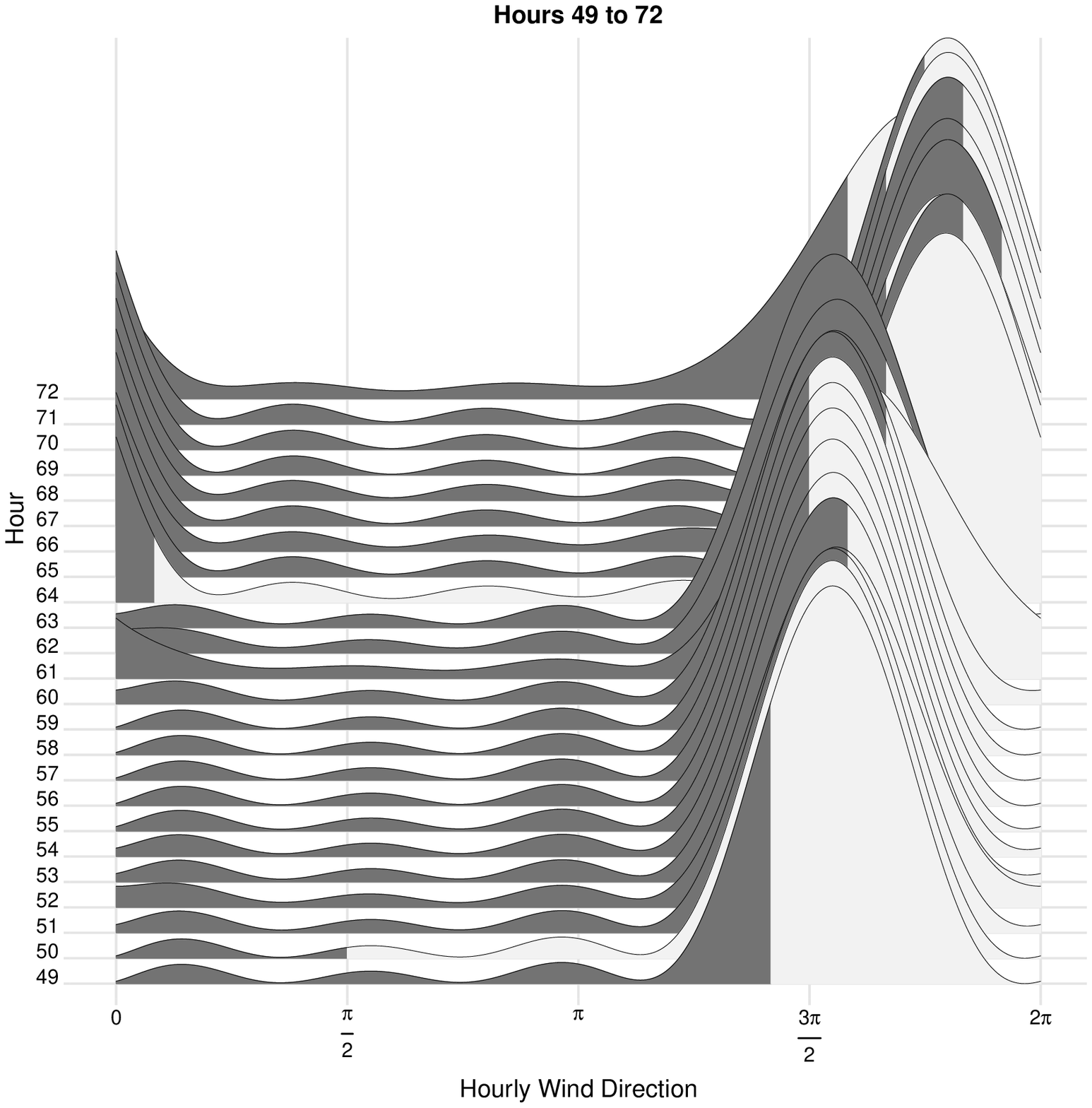}
\includegraphics[scale=.3, bb= 0 0 504 504]{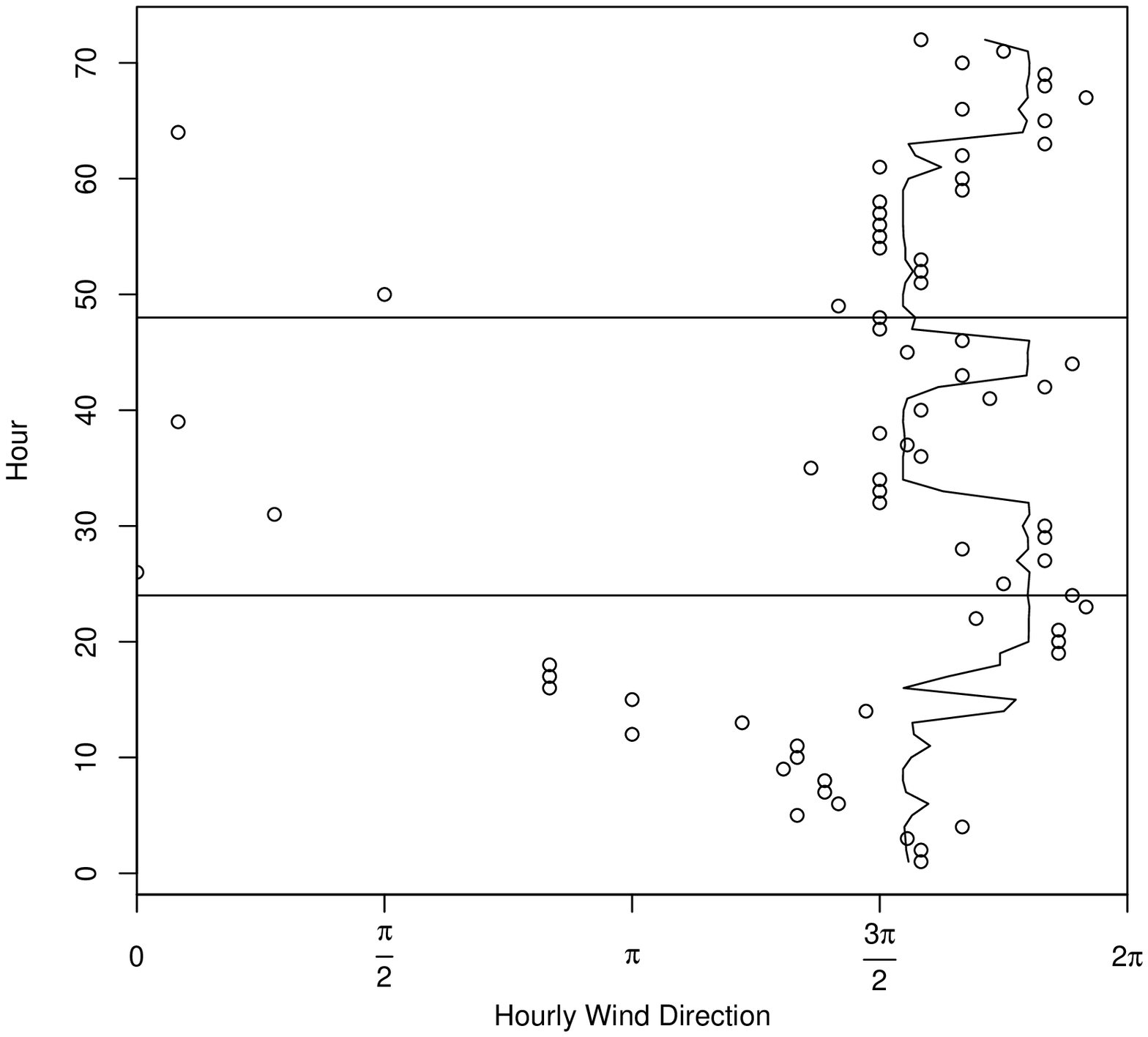}

\caption{\label{graphdatab23fisherforecasts} Wind direction data. Density forecasts for the AR(2) great-circle nonnegative trigonometric sums (NNTS) model with $M=4$ are shown. The top left plot corresponds to hours 1 to 24, the bottom left plot corresponds to hours 49 to 72, and the top right plot corresponds to hours 25 to 48. The bottom right plot shows the scatterplot of the original data with hourly wind direction in the horizontal axis and hour in the vertical axis, including the mean function (solid line). The filling grey intensity in the density forecasts changes at the corresponding observed value.}

\end{figure}

For the great-circle AR(2) NNTS model with $M=4$, Figure \ref{graphdatab23fisherforecasts} shows the density forecasts for hours 1 to 72. In addition, the scatterplot of the original data and the mean function of the model at the bottom right of Figure \ref{graphdatab23fisherforecasts} are shown to compare the density forecast with the observed
values. As can be seen, it is clear that the model presents a good fit to the data. Here, the mean function is around a mean value of 4.620 (264.707$^\circ$), which is similar to the values
reported by Fisher (1993) and Fisher and Lee (1994) for the CAR(1) and LAR(1) models. Note that there are some extreme observations, such as the observation at hour 50 (second density forecast in the bottom left plot of Figure \ref{graphdatab23fisherforecasts}), that were captured by the great-circle AR(2) NNTS model with $M=4$ because their density forecasts are multimodal. Thus, they are more difficult to capture by a unimodal model based on the von Mises distribution.

\subsection{Earthquake Occurrence and Planet Alignments}

There is debate about the prediction of the occurrence time of large earthquakes in terms of variables related to the po\-si\-tions of different astronomical bodies (e.g., the sun, moon, and planets).
In particular, the times at which two planets are in conjunction (Ip 1976; Hughes 1977; Geller 1997) have been considered as times when earthquakes are more likely to occur.

A database with the occurrence times of earthquakes of magnitude 6 (on the Richter scale) or greater from January 1, 1969 to December 15, 2019, and the right ascension and declination of the planets (excluding Pluto) has been con\-struct\-ed previously. The times of earthquakes were obtained from the NOOA Significant Earthquakes Database (National Geophysical Data Center / World Data Service (NGDC/WDS): Significant Earthquake Database. National Geophysical Data Center, NOAA. doi:10.7289/V5TD9V7K) and were trans\-formed to Julian dates. Given the number of earthquakes, right ascensions and declinations were calculated using the Skyfield (PyEphem) Python astronomical package (Rhodes 2020). The angular distance in the celestial sphere between each pair of planets was calculated from the right ascension and declination of the planets. In addition, indicator variables were calculated for the case where the angular dis\-tance between two planets was less than 2.5$^\circ$.

Here, we consider a regression model with the occurrence time (in Julian time) of earthquakes with magnitude 6 or greater, transformed into a circular variable as a dependent variable, and the angular distances between planets and in\-di\-ca\-tors of separation between planets of less than 2.5$^\circ$ were used as explanatory variables. There is a total of 1153 observations on 42 explanatory variables.
We did not find any predictive power to explain the times at which an earthquake of mag\-ni\-tude 6 or greater will occur. Here, we employed the LASSO regularization algorithm (Tibshirani 1996; Hastie et al. 2009) with a penalty alpha parameter of 0.5. For the  great-circle NNTS models shown in Table \ref{Table03earthquakes}, we found that there were variables with non-zero coefficients only for cases where $M= 1, 2, \mbox{ or } 7$. In addition, the range, Kuiper and Watson tests did not reject the circular uniformity of the PIT
values at a 1\% significance level, although there was no predictive power (the maximum $R^2$ value was 0.032), and a model with $i.i.d.$ uniform distributed occurrence times was preferable (last row of Table \ref{Table03earthquakes}). For all small-circle NNTS models (Table \ref{Table03earthquakes}),
the range, Kuiper and Watson tests rejected the null hypothesis of circular uniformity of the PIT values.
The same results were obtained by considering different values for the penalty alpha parameter. Thus, for the fitted small-circle the $\phi$ angles are very near zero and the $\alpha$ angle is near $\frac{\pi}{4}$, thereby rendering the small circle a great circle with vectors $\bm{b}$ and $\bm{a}$ with equal weights where squares are equal to 0.5.

\section{Conclusions}

NNTS models allow us to model circular variables when their densities present skewness and multimodality. Previous models only considered unimodal and symmetric models
based on the von Mises or other symmetric densities. By exploiting the fact that the parameter space of
NNTS models is a hypersphere, regression models were constructed, in which great and small circles were equivalent to regression planes in the classical linear model.
The different points on the great (small) circle specified a different NNTS density, and the significance of the explanatory variables was evaluated according to
their power to explain different angle rotations along the great (small) circle. This evaluation was performed by considering transformation of the original circular variable
into a linear variable where common models (e.g., ordinary least-squares regression or linear time series models) were applied to the transformed variable.
The proposed circular regression model generates fitted (forecast) NNTS
densities; thus, the goodness of fit of the regression model can be assessed using tests of uniformity for the probability integral transforms in consideration of the fitted (forecast) NNTS densities. The usefulness of NNTS models with skewness and multimodality was shown in the examples presented in this paper.

\section{Acknowledgements}

The authors wish to thank the Asociaci\'on Mexicana de Cultura, A.C. for its support.

\newpage
\renewcommand{\baselinestretch}{1.00}

\begin{table}[t]
\caption{\label{Table04simulationGC} Great-circle regression simulations: Beta parameter estimates, rejection rates (RR) for the null hypothesis $H_0:\beta=0$ and acceptance rates (AR) for the circular uniformity tests for the PIT (Probability Integral Transform) values for the range (R), Kuiper (K) and Watson (W) tests considering a 5\% significance level.}
\centering
\scalebox{.7}{
\begin{tabular}{|c|c|c|c|c|c|c|c|c|c|c|c|c|c|c|}
\hline
case /       & obs & $\beta_1$ & $\beta_2$ & $\beta_3$ & $\beta_4$ & $\beta_5$ & RR        & RR        & RR        & RR        & RR         & AR    & AR     & AR     \\
(eigen)      &     &           &           &           &           &           & $\beta_1$ & $\beta_2$ & $\beta_3$ & $\beta_4$ & $\beta_5$  & (R)   & (K)    & (W)    \\
vectors      &     &           &           &           &           &           &           &           &           &           &            &       &        &        \\
\hline
\hline
1  &  & 0 & 0 & 0 & 0 & 0 &  &  &  &  &  &  &  &  \\
\hline
\hline
known     & 25     & 0.165 & 0.142 & 0.191 & 0.138 & 0.172 & 0.06 & 0.04 & 0.05 & 0.06 & 0.05 & 0.96 & 0.95 & 0.96 \\
          & 50     & 0.112 & 0.100 & 0.117 & 0.104 & 0.109 & 0.06 & 0.03 & 0.05 & 0.06 & 0.07 & 0.96 & 0.94 & 0.95 \\
          & 100    & 0.077 & 0.068 & 0.075 & 0.078 & 0.074 & 0.05 & 0.04 & 0.05 & 0.06 & 0.06 & 0.96 & 0.96 & 0.96 \\
          & 200    & 0.053 & 0.052 & 0.050 & 0.053 & 0.053 & 0.05 & 0.05 & 0.03 & 0.05 & 0.06 & 0.94 & 0.96 & 0.96 \\
          & 500    & 0.034 & 0.032 & 0.031 & 0.034 & 0.034 & 0.05 & 0.05 & 0.05 & 0.04 & 0.05 & 0.96 & 0.94 & 0.95 \\
          & 1000   & 0.024 & 0.023 & 0.021 & 0.024 & 0.024 & 0.04 & 0.06 & 0.05 & 0.04 & 0.05 & 0.95 & 0.94 & 0.95 \\
\hline
est.      & 25     & 0.784 & 0.661 & 0.881 & 0.654 & 0.940 & 0.06 & 0.06 & 0.05 & 0.04 & 0.07 & 0.95 & 0.93 & 0.94 \\
          & 50     & 0.326 & 0.275 & 0.429 & 0.243 & 0.202 & 0.04 & 0.04 & 0.06 & 0.05 & 0.08 & 0.97 & 0.95 & 0.96 \\
          & 100    & 0.109 & 0.101 & 0.101 & 0.105 & 0.115 & 0.05 & 0.05 & 0.04 & 0.06 & 0.06 & 0.95 & 0.93 & 0.93 \\
          & 200    & 0.064 & 0.062 & 0.061 & 0.063 & 0.068 & 0.04 & 0.04 & 0.06 & 0.05 & 0.07 & 0.92 & 0.80 & 0.82 \\
          & 500    & 0.039 & 0.036 & 0.034 & 0.038 & 0.038 & 0.06 & 0.05 & 0.05 & 0.05 & 0.04 & 0.92 & 0.59 & 0.66 \\
          & 1000   & 0.027 & 0.025 & 0.024 & 0.026 & 0.027 & 0.04 & 0.04 & 0.04 & 0.05 & 0.05 & 0.89 & 0.43 & 0.48 \\
\hline
\hline
2  &  & 0.3 & 0.2 & 0.15 & 0.2 & 0.3 &  &  &  &  &  &  &  &  \\
\hline
\hline
known     & 25     & 0.213 & 0.200 & 0.205 & 0.159 & 0.210 & 0.16 & 0.18 & 0.07 & 0.12 & 0.14 & 0.95 & 0.94 & 0.95 \\
          & 50     & 0.209 & 0.174 & 0.129 & 0.139 & 0.200 & 0.33 & 0.30 & 0.09 & 0.17 & 0.37 & 0.94 & 0.94 & 0.92 \\
          & 100    & 0.205 & 0.147 & 0.110 & 0.140 & 0.201 & 0.64 & 0.40 & 0.21 & 0.35 & 0.63 & 0.95 & 0.93 & 0.94 \\
          & 200    & 0.194 & 0.117 & 0.087 & 0.118 & 0.178 & 0.89 & 0.51 & 0.28 & 0.51 & 0.82 & 0.93 & 0.92 & 0.93 \\
          & 500    & 0.207 & 0.127 & 0.091 & 0.127 & 0.193 & 1    & 0.91 & 0.70 & 0.89 & 1    & 0.94 & 0.90 & 0.89 \\
          & 1000   & 0.209 & 0.131 & 0.093 & 0.129 & 0.198 & 1    & 1    & 0.92 & 0.98 & 1    & 0.95 & 0.85 & 0.85 \\
\hline
est.      & 25     & 0.752 & 0.748 & 0.772 & 0.543 & 0.878 & 0.14 & 0.09 & 0.08 & 0.1  & 0.14 & 0.95 & 0.92 & 0.92 \\
          & 50     & 0.315 & 0.255 & 0.259 & 0.223 & 0.274 & 0.27 & 0.21 & 0.07 & 0.12 & 0.23 & 0.95 & 0.97 & 0.98 \\
          & 100    & 0.250 & 0.174 & 0.156 & 0.172 & 0.223 & 0.57 & 0.34 & 0.17 & 0.28 & 0.53 & 0.96 & 0.95 & 0.96 \\
          & 200    & 0.209 & 0.125 & 0.097 & 0.130 & 0.186 & 0.84 & 0.47 & 0.25 & 0.47 & 0.76 & 0.94 & 0.86 & 0.89 \\
          & 500    & 0.220 & 0.132 & 0.098 & 0.135 & 0.200 & 0.99 & 0.87 & 0.69 & 0.88 & 0.98 & 0.92 & 0.64 & 0.70 \\
          & 1000   & 0.221 & 0.136 & 0.099 & 0.133 & 0.201 & 1    & 0.98 & 0.91 & 0.97 & 1    & 0.92 & 0.53 & 0.57 \\
\hline
\hline
3  &  & 0 & 0 & 0 & 0 & 0.3 &  &  &  &  &  &  &  &  \\
\hline
\hline
known     & 25     & 0.159 & 0.131 & 0.186 & 0.133 & 0.313 & 0.06 & 0.05 & 0.05 & 0.05 & 0.30 & 0.96 & 0.96 & 0.95 \\
          & 50     & 0.102 & 0.095 & 0.109 & 0.097 & 0.272 & 0.05 & 0.04 & 0.04 & 0.04 & 0.58 & 0.96 & 0.94 & 0.94 \\
          & 100    & 0.071 & 0.067 & 0.076 & 0.069 & 0.268 & 0.05 & 0.05 & 0.05 & 0.05 & 0.86 & 0.94 & 0.95 & 0.95 \\
          & 200    & 0.050 & 0.045 & 0.051 & 0.048 & 0.265 & 0.04 & 0.04 & 0.06 & 0.05 & 0.98 & 0.95 & 0.94 & 0.94 \\
          & 500    & 0.032 & 0.031 & 0.031 & 0.031 & 0.265 & 0.06 & 0.05 & 0.05 & 0.04 & 1    & 0.96 & 0.93 & 0.94 \\
          & 1000   & 0.022 & 0.022 & 0.023 & 0.023 & 0.267 & 0.05 & 0.04 & 0.05 & 0.04 & 1    & 0.94 & 0.95 & 0.95 \\
\hline
est.      & 25     & 0.484 & 0.347 & 0.641 & 0.348 & 0.488 & 0.07 & 0.06 & 0.06 & 0.04 & 0.13 & 0.95 & 0.94 & 0.94 \\
          & 50     & 0.231 & 0.167 & 0.223 & 0.256 & 0.392 & 0.05 & 0.05 & 0.07 & 0.04 & 0.40 & 0.96 & 0.96 & 0.96 \\
          & 100    & 0.099 & 0.094 & 0.104 & 0.103 & 0.287 & 0.06 & 0.05 & 0.05 & 0.05 & 0.73 & 0.93 & 0.92 & 0.93 \\
          & 200    & 0.059 & 0.052 & 0.063 & 0.058 & 0.276 & 0.04 & 0.04 & 0.07 & 0.05 & 0.93 & 0.93 & 0.80 & 0.83 \\
          & 500    & 0.035 & 0.035 & 0.036 & 0.035 & 0.278 & 0.06 & 0.04 & 0.06 & 0.04 & 1    & 0.92 & 0.59 & 0.67 \\
          & 1000   & 0.024 & 0.025 & 0.025 & 0.026 & 0.279 & 0.06 & 0.05 & 0.05 & 0.03 & 1    & 0.89 & 0.49 & 0.50 \\
\hline
4  &  &  &  &  &  & 0.3 &  &  &  &  &  &  &  & \\
\hline
known     & 25     &       &       &       &       & 0.298 &      &      &      &      & 0.36 & 0.96 & 0.96 & 0.96 \\
          & 50     &       &       &       &       & 0.272 &      &      &      &      & 0.62 & 0.96 & 0.94 & 0.95 \\
          & 100    &       &       &       &       & 0.268 &      &      &      &      & 0.86 & 0.96 & 0.95 & 0.95 \\
          & 200    &       &       &       &       & 0.265 &      &      &      &      & 0.98 & 0.94 & 0.95 & 0.96 \\
          & 500    &       &       &       &       & 0.265 &      &      &      &      & 1    & 0.95 & 0.94 & 0.94 \\
          & 1000   &       &       &       &       & 0.267 &      &      &      &      & 1    & 0.94 & 0.95 & 0.95 \\
\hline
est.      & 25     &       &       &       &       & 0.644 &      &      &      &      & 0.19 & 0.95 & 0.95 & 0.96 \\
          & 50     &       &       &       &       & 0.399 &      &      &      &      & 0.43 & 0.97 & 0.96 & 0.97 \\
          & 100    &       &       &       &       & 0.285 &      &      &      &      & 0.73 & 0.94 & 0.92 & 0.93 \\
          & 200    &       &       &       &       & 0.277 &      &      &      &      & 0.93 & 0.94 & 0.79 & 0.82 \\
          & 500    &       &       &       &       & 0.277 &      &      &      &      & 1    & 0.93 & 0.59 & 0.66 \\
          & 1000   &       &       &       &       & 0.279 &      &      &      &      & 1    & 0.89 & 0.49 & 0.50 \\
\hline
\end{tabular}}
\renewcommand{\baselinestretch}{1}
\end{table}

\newpage
\renewcommand{\baselinestretch}{1.00}

\begin{table}[t]
\caption{\label{Table05imulationSC} Small-circle regression simulations: Beta parameter estimates, rejection rates (RR) for the null hypothesis $H_0:\beta=0$ and acceptance rates (AR) for the
circular uniformity tests for the PIT (Probability Integral Transform) values for the range, Kuiper and Watson tests considering a 5\% significance level.}
\centering
\scalebox{.7}{
\begin{tabular}{|c|c|c|c|c|c|c|c|c|c|c|c|c|c|c|}
\hline
case /       & obs & $\beta_1$ & $\beta_2$ & $\beta_3$ & $\beta_4$ & $\beta_5$ & RR        & RR        & RR        & RR        & RR         & AR    & AR     & AR     \\
(eigen)      &     &           &           &           &           &           & $\beta_1$ & $\beta_2$ & $\beta_3$ & $\beta_4$ & $\beta_5$  & (R)   & (K)    & (W)    \\
vectors      &     &           &           &           &           &           &           &           &           &           &            &       &        &        \\
\hline
\hline
1  &  & 0 & 0 & 0 & 0 & 0 &  &  &  &  &  &  &  &   \\
\hline
\hline
known     & 25   & 0.177 & 0.146 & 0.214 & 0.144 & 0.176 & 0.06 & 0.05 & 0.06 & 0.06 & 0.05 & 0.96 & 0.96 & 0.97 \\
          & 50   & 0.113 & 0.101 & 0.130 & 0.111 & 0.107 & 0.03 & 0.05 & 0.05 & 0.05 & 0.05 & 0.97 & 0.96 & 0.95 \\
          & 100  & 0.075 & 0.078 & 0.079 & 0.080 & 0.074 & 0.04 & 0.07 & 0.06 & 0.06 & 0.05 & 0.92 & 0.95 & 0.95 \\
          & 200  & 0.053 & 0.054 & 0.057 & 0.058 & 0.052 & 0.03 & 0.05 & 0.07 & 0.04 & 0.07 & 0.94 & 0.95 & 0.95 \\
          & 500  & 0.035 & 0.032 & 0.035 & 0.036 & 0.035 & 0.04 & 0.04 & 0.06 & 0.05 & 0.05 & 0.95 & 0.95 & 0.94 \\
          & 1000 & 0.024 & 0.025 & 0.025 & 0.026 & 0.025 & 0.04 & 0.05 & 0.05 & 0.06 & 0.04 & 0.92 & 0.95 & 0.94 \\
\hline
est.      & 25   & 1.009 & 0.669 & 1.591 & 0.657 & 1.319 & 0.07 & 0.06 & 0.07 & 0.05 & 0.14 & 0.92 & 0.87 & 0.87 \\
          & 50   & 0.203 & 0.183 & 0.275 & 0.212 & 0.267 & 0.04 & 0.04 & 0.07 & 0.05 & 0.09 & 0.91 & 0.67 & 0.67 \\
          & 100  & 0.092 & 0.093 & 0.090 & 0.091 & 0.096 & 0.05 & 0.08 & 0.06 & 0.04 & 0.07 & 0.84 & 0.36 & 0.33 \\
          & 200  & 0.050 & 0.053 & 0.057 & 0.055 & 0.063 & 0.04 & 0.06 & 0.07 & 0.04 & 0.08 & 0.75 & 0.16 & 0.18 \\
          & 500  & 0.033 & 0.032 & 0.033 & 0.033 & 0.038 & 0.04 & 0.05 & 0.05 & 0.05 & 0.08 & 0.64 & 0.02 & 0.02 \\
          & 1000 & 0.022 & 0.023 & 0.023 & 0.024 & 0.028 & 0.04 & 0.05 & 0.05 & 0.06 & 0.09 & 0.61 & 0    & 0    \\
\hline
\hline
2  &  & 0.3 & 0.2 & 0.15 & 0.2 & 0.3 &  &  &  &  &  &  &  &  \\
\hline
\hline
known     & 25   & 0.222 & 0.219 & 0.226 & 0.171 & 0.245 & 0.18 & 0.22 & 0.08 & 0.14 & 0.17 & 0.97 & 0.96 & 0.96 \\
          & 50   & 0.237 & 0.192 & 0.154 & 0.157 & 0.240 & 0.39 & 0.34 & 0.12 & 0.20 & 0.46 & 0.95 & 0.95 & 0.95 \\
          & 100  & 0.233 & 0.171 & 0.123 & 0.160 & 0.242 & 0.69 & 0.47 & 0.25 & 0.40 & 0.77 & 0.94 & 0.96 & 0.96 \\
          & 200  & 0.228 & 0.136 & 0.101 & 0.137 & 0.218 & 0.94 & 0.61 & 0.34 & 0.57 & 0.93 & 0.94 & 0.94 & 0.94 \\
          & 500  & 0.239 & 0.154 & 0.107 & 0.149 & 0.230 & 1    & 0.97 & 0.80 & 0.97 & 1    & 0.95 & 0.90 & 0.90 \\
          & 1000 & 0.243 & 0.158 & 0.110 & 0.152 & 0.235 & 1    & 1    & 0.96 & 1    & 1    & 0.95 & 0.84 & 0.84 \\
\hline
est.      & 25   & 1.289 & 0.580 & 0.808 & 0.804 & 1.133 & 0.19 & 0.13 & 0.07 & 0.07 & 0.17 & 0.93 & 0.88 & 0.86 \\
          & 50   & 0.284 & 0.197 & 0.187 & 0.186 & 0.222 & 0.29 & 0.16 & 0.09 & 0.10 & 0.25 & 0.88 & 0.72 & 0.68 \\
          & 100  & 0.207 & 0.149 & 0.114 & 0.161 & 0.216 & 0.50 & 0.32 & 0.16 & 0.30 & 0.53 & 0.84 & 0.43 & 0.42 \\
          & 200  & 0.183 & 0.114 & 0.088 & 0.110 & 0.172 & 0.76 & 0.46 & 0.26 & 0.44 & 0.75 & 0.79 & 0.19 & 0.19 \\
          & 500  & 0.192 & 0.124 & 0.087 & 0.122 & 0.187 & 0.95 & 0.87 & 0.70 & 0.88 & 0.96 & 0.74 & 0.03 & 0.02 \\
          & 1000 & 0.200 & 0.128 & 0.091 & 0.123 & 0.190 & 1    & 0.98 & 0.90 & 0.98 & 1    & 0.69 & 0.01 & 0    \\
\hline
\hline
3  &  & 0 & 0 & 0 & 0 & 0.3 &  &  &  &  &  &  &  &  \\
\hline
\hline
known     & 25   & 0.166 & 0.142 & 0.192 & 0.135 & 0.368 & 0.07 & 0.06 & 0.03 & 0.05 & 0.37 & 0.96 & 0.95 & 0.95 \\
          & 50   & 0.113 & 0.099 & 0.123 & 0.107 & 0.326 & 0.07 & 0.04 & 0.06 & 0.05 & 0.69 & 0.95 & 0.95 & 0.94 \\
          & 100  & 0.074 & 0.069 & 0.075 & 0.075 & 0.316 & 0.04 & 0.04 & 0.04 & 0.04 & 0.93 & 0.94 & 0.94 & 0.95 \\
          & 200  & 0.053 & 0.049 & 0.052 & 0.050 & 0.312 & 0.07 & 0.05 & 0.06 & 0.04 & 1    & 0.95 & 0.95 & 0.96 \\
          & 500  & 0.033 & 0.031 & 0.032 & 0.032 & 0.309 & 0.04 & 0.05 & 0.04 & 0.04 & 1    & 0.95 & 0.96 & 0.94 \\
          & 1000 & 0.022 & 0.022 & 0.023 & 0.022 & 0.310 & 0.04 & 0.04 & 0.04 & 0.04 & 1    & 0.97 & 0.95 & 0.95 \\
\hline
est.      & 25   & 0.754 & 0.577 & 0.703 & 0.501 & 0.595 & 0.11 & 0.09 & 0.07 & 0.04 & 0.15 & 0.93 & 0.91 & 0.91 \\
          & 50   & 0.186 & 0.166 & 0.191 & 0.177 & 0.267 & 0.06 & 0.04 & 0.06 & 0.04 & 0.32 & 0.89 & 0.72 & 0.69 \\
          & 100  & 0.091 & 0.090 & 0.084 & 0.079 & 0.243 & 0.06 & 0.06 & 0.04 & 0.04 & 0.64 & 0.84 & 0.38 & 0.36 \\
          & 200  & 0.056 & 0.056 & 0.055 & 0.051 & 0.239 & 0.05 & 0.04 & 0.05 & 0.04 & 0.88 & 0.76 & 0.18 & 0.19 \\
          & 500  & 0.030 & 0.029 & 0.031 & 0.030 & 0.250 & 0.04 & 0.04 & 0.07 & 0.04 & 0.98 & 0.72 & 0.04 & 0.04 \\
          & 1000 & 0.020 & 0.020 & 0.021 & 0.020 & 0.254 & 0.05 & 0.04 & 0.05 & 0.04 & 0.99 & 0.67 & 0    & 0    \\
\hline
4  &  &  &  &  &  & 0.3 &  &  &  &  &  &  &  &  \\
\hline
known     & 25   &       &       &       &       & 0.361 &      &      &      &      & 0.44 & 0.96 & 0.97 & 0.96 \\
          & 50   &       &       &       &       & 0.324 &      &      &      &      & 0.70 & 0.95 & 0.97 & 0.96 \\
          & 100  &       &       &       &       & 0.316 &      &      &      &      & 0.93 & 0.93 & 0.95 & 0.96 \\
          & 200  &       &       &       &       & 0.312 &      &      &      &      & 1    & 0.96 & 0.95 & 0.96 \\
          & 500  &       &       &       &       & 0.308 &      &      &      &      & 1    & 0.95 & 0.95 & 0.95 \\
          & 1000 &       &       &       &       & 0.309 &      &      &      &      & 1    & 0.95 & 0.95 & 0.95 \\
\hline
est.      & 25   &       &       &       &       & 0.626 &      &      &      &      & 0.18 & 0.91 & 0.89 & 0.89 \\
          & 50   &       &       &       &       & 0.268 &      &      &      &      & 0.34 & 0.88 & 0.60 & 0.57 \\
          & 100  &       &       &       &       & 0.242 &      &      &      &      & 0.63 & 0.84 & 0.35 & 0.32 \\
          & 200  &       &       &       &       & 0.238 &      &      &      &      & 0.88 & 0.77 & 0.16 & 0.17 \\
          & 500  &       &       &       &       & 0.250 &      &      &      &      & 0.98 & 0.70 & 0.03 & 0.04 \\
          & 1000 &       &       &       &       & 0.254 &      &      &      &      & 0.99 & 0.67 & 0    & 0    \\
\hline
\end{tabular}}
\renewcommand{\baselinestretch}{1}
\end{table}

\newpage
\renewcommand{\baselinestretch}{1.00}

\begin{table}
	\caption{\label{Table01blueperiwinkles} Blue periwinkle data. Coefficients of determination, p-values of the range, Kuiper and Watson circular uniformity tests for PIT values, and loglikelihood for the great- and small-circle nonnegative trigonometric series (NNTS) models, Fisher and Lee (1992) mean and mixed models, Presnell et al. (1998) $SPML$ mixed model, and Rivest et al. (2015) mixed models are shown. These models include the $distance$ travelled by small periwinkles as the explanatory variable.}
\centering
\scalebox{.7}{
\begin{tabular}{|c|c|c|c|c|c|c|c|}
\hline
\multicolumn{8}{|c|}{Great-circle model $Y = \beta_1(distance - 27)I(distance<27) +  error$} \\
\hline
  $M$ ($\alpha$ angle)  & $distance$ coef.           & $R^2_{cos}$ & $R^2$ & Range  & Kuiper  &  Watson  & loglik  \\
                        & (std. error, p-value)      &             &       & p-value& p-value &  p-value &          \\
\hline
1 &  0.042 (0.014, 0.006)  & 0.847 & 0.227     & 0.062       & $<$0.01        &  $<$0.01       & -39.643              \\
2 &  0.230 (0.064, 0.001)  & 0.747 & 0.302     & 0.024       & $<$0.01        &  $<$0.01       & -32.135              \\
3 &  0.043 (0.033, 0.202)  & 0.618 & 0.054     & 0.003       & (0.10,0.15)    &  $>$0.10       & -35.406              \\
4 &  0.032 (0.027, 0.245)  & 0.508 & 0.045     & 0.070       & $>$0.15        &  $>$0.10       & -33.625              \\
5 &  0.028 (0.023, 0.241)  & 0.451 & 0.045     & 0.451       & $>$0.15        &  $>$0.10       & -34.178              \\
6 &  0.015 (0.027, 0.582)  & 0.395 & 0.010     & 0.434       & $>$0.15        &  $>$0.10       & -33.506              \\
7 &  0.032 (0.037, 0.392)  & 0.337 & 0.024     & 0.154       & (0.05,0.10)    &  $>$0.10       & -34.881              \\
8 &  0.122 (0.093, 0.200)  & 0.305 & 0.054     & 0.054       & $<$0.01        & (0.01,0.025)   & -43.308              \\
\hline
\multicolumn{8}{|c|}{Small-circle model $Y = \beta_1(distance - 27)I(distance<27) +  error$} \\
\hline
1 (0.332)   & 0.006 (0.289, 0.984)   & 0.897    & 0.000     & 0.000    &  $<$0.01       & $<$0.01        & -38.042        \\
2 (0.464)   & 0.008 (0.073, 0.912)   & 0.724    & 0.000     & 0.005    &  $<$0.01       & $<$0.01        & -32.974        \\
3 (0.566)   & 0.033 (0.024, 0.175)   & 0.758    & 0.060     & 0.003    &  $<$0.01       & $<$0.01        & -25.056        \\
4 (0.641)   & 0.017 (0.033, 0.608)   & 0.698    & 0.009     & 0.053    &  $<$0.01       & $<$0.01        & -23.077        \\
5 (0.711)   & 0.321 (0.055, 0.000)   & 0.615    & 0.534     & 0.054    &  $<$0.01       & $<$0.01        & -16.902        \\
6 (0.779)   & 0.136 (0.117, 0.254)   & 0.565    & 0.043     & 0.115    & (0.025,0.05)   & (0.025,0.05)   & -19.176        \\
7 (0.835)   & 0.049 (0.136, 0.722)   & 0.504    & 0.004     & 0.101    &  $>$0.15       & $>$0.10        & -25.394        \\
8 (0.884)   & 0.300 (0.089, 0.002)   & 0.440    & 0.276     & 0.329    &  $>$0.15       & $>$0.10        & -19.786        \\
\hline
\multicolumn{8}{|c|}{$i.i.d.$ Uniforms} \\
\hline
0 &                        &          &           &  0.000      &    $<$0.01    &  $<$0.01     &  -56.974                  \\
\hline
\multicolumn{2}{|c|}{Parameter}              & & & Range   & Kuiper  & Watson  &  loglik  \\
\multicolumn{2}{|c|}{estimates (std. error)} & & & p-value & p-value & p-value &          \\
\hline
\multicolumn{8}{|c|}{Fisher and Lee (1992) Mean von Mises $vM(\mu,\kappa)$ model} \\
\multicolumn{8}{|c|}{$\mu =  \mu_0 + \arctan(\beta (distance - \overline{distance}))$} \\
\hline
$\hat{\mu_0}$=1.694 (0.112)  & $\hat{\beta}$= -0.0065 (0.0022)& & & 0.446 &  $>$0.15      & $>$0.10      &  -29.452  \\
$\hat{\kappa}$=3.203 (0.707) &                                & & &       &               &              &           \\
\hline
\multicolumn{8}{|c|}{Fisher and Lee (1992) Mixed von Mises $vM(\mu,\kappa)$ model} \\
\multicolumn{8}{|c|}{$\mu =  \mu_0 + \arctan(\beta distance)$ $\kappa =  e^{\gamma_0 + \gamma_1 (distance - \overline{distance})}$} \\
\hline
$\hat{\mu_0}$=2.034 (0.190)    & $\hat{\beta}$= -0.0045 (0.0012)& & & 0.248 & $>$0.15      & $>$0.10      &  -18.963  \\
$\hat{\gamma}_0$=1.785 (0.248) & $\hat{\gamma}_1$=0.045 (0.010) & & &       &              &              &           \\
\hline
\multicolumn{8}{|c|}{Presnell, Morrison and Little (1998) Mixed $SPML$ angular normal $AN(\mu,\gamma))$ model} \\
\multicolumn{8}{|c|}{(reported estimates)} \\
\multicolumn{8}{|c|}{$\mu = \left[\arctan\left(\frac{\beta_{0s}+\beta_{1s}distance}{\beta_{0c}+\beta_{1c}distance} \right) + \pi I(\beta_{0c} + \beta_{1c}distance)\right] \mod 2\pi$} \\
\multicolumn{8}{|c|}{$\gamma=\sqrt{(\beta_{0s}+\beta_{1s}distance)^2 + (\beta_{0c}+\beta_{1c}distance)^2}$} \\
\hline
$\hat{\beta}_{0c}$=-1.228 (0.423) & $\hat{\beta}_{1c}$=0.030 (0.008)  & & & 0.134 & $>$0.15      & $>$0.10      &  -20.601  \\
$\hat{\beta}_{0s}$= 0.157 (0.451) & $\hat{\beta}_{1s}$=0.049 (0.012)  & & &       &              &              &           \\
\hline
\multicolumn{8}{|c|}{Rivest et al. (2016) Mixed von Mises $vM(\mu,\kappa)$ full model (reported estimates)} \\
\multicolumn{8}{|c|}{$\mu=\arctan\left(\frac{\beta_{0s}+\beta_{1s}distance}{1 + \beta_{1c}distance} \right)$ $\kappa=\gamma distance$} \\
\hline
$\hat{\beta}_{0s}$= -0.062 (0.440)& $\hat{\beta}_{1c}$= 0.026 (0.006) & & & 0.000 & $<$0.01      & $<$0.01      &  -21.817  \\
$\hat{\gamma}$=0.134              & $\hat{\beta}_{1s}$= 0.060 (0.028) & & &       &              &              &           \\
\hline
\multicolumn{8}{|c|}{Rivest et al. (2016) Mixed von Mises $vM(\mu,\kappa)$ reciprocal model (reported estimates)} \\
\multicolumn{8}{|c|}{$\mu=\arctan\left(\frac{1}{\beta_{0c}+\beta_{1c}distance} \right)$ $\kappa=\gamma distance$} \\
\hline
$\hat{\beta}_{0c}$= -0.790 (0.390)& $\hat{\beta}_{1c}$= 0.014 (0.006) & & & 0.000 & $<$0.01      & (0.01,0.025) &  -22.312  \\
$\hat{\gamma}$=0.129              &                                   & & &       &              &              &           \\
\hline
\end{tabular}}
\renewcommand{\baselinestretch}{1}
\end{table}

\newpage
\renewcommand{\baselinestretch}{1.00}
\begin{table}
	\caption{\label{Table02winddirections1to72AR1} Wind direction data. Results for the AR(1) great and small-circle NNTS models and CAR(1) model for 72 observations are shown. The autoregressive parameter estimates, coefficient of determination, p-values of the range, Kuiper and Watson tests, and loglikelihood values are shown. The last rows of the table present the results for the AR(2) great-circle nonnegative trigonometric series (NNTS) final model with $M=4$. The other parameter estimates of the CAR(1) are equal to $\hat{\mu}=$5.1617 (0.2675) and $\hat{\kappa}=$2.4493 (0.3415).}
\centering
\scalebox{.7}{
\begin{tabular}{|c|c|c|c|c|c|c|c|}
\hline
\multicolumn{8}{|c|}{Great-circle models} \\
\hline
  $M$ & AR(1) Coeff. &  $R^2_{cos}$ & Range    & Kuiper  & Watson  & loglik & $\alpha$ angle \\
      & (std. error) &              & p-value  & p-value & p-value &        &                \\
\hline
1 & 0.6222 (0.0900)                   & 0.837 & 0.000 & $<$0.01       & $<$0.01 &  -96.316 & \\
\hline
2 & 0.5041 (0.1002)                   & 0.701 & 0.146 & $<$0.01       & $<$0.01 &  -83.732 & \\
\hline
3 & 0.5332 (0.0979)                   & 0.571 & 0.120 & $>$0.15       & $>$0.10 &  -87.123 & \\
\hline
4 & 0.5276 (0.0984)                   & 0.462 & 0.183 & $>$0.15       & $>$0.10 &  -78.050 & \\
\hline
5 & 0.4478 (0.1039)                   & 0.384 & 0.009 & (0.05,0.10)   & $>$0.10 &  -86.861 & \\
\hline
\multicolumn{8}{|c|}{Small-circle models} \\
\hline
\hline
1 &  0.3430 (0.1133)                  & 0.875 & 0.073 & $<$0.01       & $<$0.01 &  -90.682 & 0.411 \\
\hline
2 &  0.0012 (0.1172)                  & 0.775 & 0.006 & $<$0.01       & $<$0.01 &  -77.201 & 0.596 \\
\hline
3 &  0.1377 (0.1169)                  & 0.651 & 0.088 & $<$0.01       & $<$0.01 &  -66.531 & 0.724 \\
\hline
4 & -0.0715 (0.1168)                  & 0.474 & 0.001 & $<$0.01       & $<$0.01 &  -75.849 & 0.806 \\
\hline
5 &  0.1778 (0.1152)                  & 0.338 & 0.001 & $<$0.01       & $<$0.01 &  -91.383 & 0.849 \\
\hline
\multicolumn{8}{|c|}{$i.i.d.$ Uniforms} \\
\hline
0 &                                   &       & 0.000 & $<$0.01       & $<$0.01 &  -132.327 &  \\
\hline
\hline
\multicolumn{8}{|c|}{Fisher and Lee (1993 and 1994) $CAR(1)$ von Mises $vM(\mu_t,\kappa)$ model} \\
\multicolumn{8}{|c|}{(reported estimates)} \\
\multicolumn{8}{|c|}{$\mu_t = \mu + 2 \arctan (\alpha \tan(0.5(\theta_{t-1}-\mu)))$} \\
\hline
  & 0.6695 (0.1492)                   &       & 0.106 &  $<$0.01      & $<$0.01  &  -81.370 &  \\
\hline
\multicolumn{8}{|c|}{Great-circle final model} \\
\hline
  $M$ & AR Coeffs.             &  $R^2_{cos}$ & Range   & Kuiper  & Watson  &  loglik & $\alpha$ angle \\
      & (std. error)           &              & p-value & p-value & p-value &         & \\
\hline
4     & AR1: 0.2988 (0.1047)                  & 0.497        & 0.295 & $>$0.15  & $>$0.10   &  -73.967 &  \\
      & AR2: 0.4225 (0.1049)                  &              &       &          &         &          &  \\
\hline
\end{tabular}}
\renewcommand{\baselinestretch}{1}
\end{table}

\newpage
\renewcommand{\baselinestretch}{1.00}
\begin{table}
	\caption{\label{Table03earthquakes} Earthquake data. The number of variables with non-zero parameter estimates (r), p-values of Kuiper and Watson tests, and loglikelihoods for the great- and small-circle nonnegative trigonometric sums (NNTS) models when applying LASSO procedures with penalty coefficients $\lambda_{min}$ and $\lambda_{1se}$ are shown.}
\centering
\scalebox{.7}{
\begin{tabular}{|c|c|c|c|c|c|c|c|c|c|c|c|}
\hline
\multicolumn{6}{|c|}{Great-circle models $\lambda_{min}$} & \multicolumn{6}{|c|}{Small-circle models $\lambda_{min}$} \\
\hline
  $M$ & r    & Range   & Kuiper  & Watson  & loglik & r    & Range   & Kuiper  & Watson  & loglik & $\alpha$ angle \\
      &      & p-value & p-value & p-value &        &      & p-value & p-value & p-value &        &               \\
\hline
1  & 6  & 0.237 & $>$0.15  & $>$0.10  &  -2113.085 & 6  & 0.000 & $<$0.01  & $<$0.01  &  -1570.802 & 0.785 \\
2  & 6  & 0.160 & $>$0.15  & $>$0.10  &  -2111.261 & 6  & 0.000 & $<$0.01  & $<$0.01  &  -1704.455 & 0.794 \\
3  & 0  & 0.041 & $>$0.15  & $>$0.10  &  -2118.975 & 0  & 0.000 & $<$0.01  & $<$0.01  &  -1747.595 & 0.801 \\
4  & 0  & 0.049 & $>$0.15  & $>$0.10  &  -2118.944 & 0  & 0.000 & $<$0.01  & $<$0.01  &  -1830.444 & 0.811 \\
5  & 0  & 0.051 & $>$0.15  & $>$0.10  &  -2118.423 & 0  & 0.000 & $<$0.01  & $<$0.01  &  -1975.683 & 0.814 \\
6  & 0  & 0.130 & $>$0.15  & $>$0.10  &  -2118.344 & 0  & 0.000 & $<$0.01  & $<$0.01  &  -1919.651 & 0.821 \\
7  & 8  & 0.635 & $>$0.15  & $>$0.10  &  -2110.981 & 8  & 0.000 & $<$0.01  & $<$0.01  &  -2052.423 & 0.824 \\
8  & 0  & 0.258 & $>$0.15  & $>$0.10  &  -2119.634 & 0  & 0.000 & $<$0.01  & $<$0.01  &  -1974.984 & 0.831 \\
\hline
\multicolumn{6}{|c|}{Great-circle models $\lambda_{1se}$} &  \multicolumn{6}{|c|}{Small-circle models $\lambda_{1se}$} \\
\hline
1  & 0  & 0.237 & $>$0.15  & $>$0.10  &  -2119.060 & 0  & 0.000 & $<$0.01  & $<$0.01  &  -1568.112 & 0.785 \\
2  & 0  & 0.213 & $>$0.15  & $>$0.10  &  -2118.976 & 0  & 0.000 & $<$0.01  & $<$0.01  &  -1699.217 & 0.794 \\
3  & 0  & 0.041 & $>$0.15  & $>$0.10  &  -2118.975 & 0  & 0.000 & $<$0.01  & $<$0.01  &  -1747.595 & 0.801 \\
4  & 0  & 0.049 & $>$0.15  & $>$0.10  &  -2118.944 & 0  & 0.000 & $<$0.01  & $<$0.01  &  -1830.444 & 0.811 \\
5  & 0  & 0.051 & $>$0.15  & $>$0.10  &  -2118.423 & 0  & 0.000 & $<$0.01  & $<$0.01  &  -1975.683 & 0.814 \\
6  & 0  & 0.130 & $>$0.15  & $>$0.10  &  -2118.344 & 0  & 0.000 & $<$0.01  & $<$0.01  &  -1919.651 & 0.821 \\
7  & 0  & 0.281 & $>$0.15  & $>$0.10  &  -2119.345 & 0  & 0.000 & $<$0.01  & $<$0.01  &  -2025.771 & 0.824 \\
8  & 0  & 0.258 & $>$0.15  & $>$0.10  &  -2119.634 & 0  & 0.000 & $<$0.01  & $<$0.01  &  -1974.984 & 0.831 \\
\hline
\multicolumn{12}{|c|}{$i.i.d.$ Uniforms} \\
\hline
0  & 0  & 0.221 & $>$0.15  & $>$0.10  &  -2120.910 & & & & & & \\
\hline
\end{tabular}}
\renewcommand{\baselinestretch}{1}
\end{table}

\renewcommand{\baselinestretch}{1}

\newpage

\newpage

\end{document}